\documentclass[aps,showpacs,twocolumn,amsmath,amssymb]{revtex4-1}
\usepackage{graphicx}
\usepackage{epstopdf}
\usepackage{bm}
\usepackage{subfigure}
\usepackage{sidecap}

\usepackage{graphicx}
\usepackage{color}
\usepackage{epstopdf}
\usepackage{amssymb}
\usepackage{amsmath}

\usepackage{bm}
\graphicspath{{Figures/}}
\usepackage{subfigure}
\usepackage{sidecap}
\usepackage {times}

\newcommand{\si}{\sigma}
\newcommand{\al}{\alpha}
\newcommand{\bt}{\beta}

\newcommand{\lam}{\lambda}

\newcommand{\De}{\Delta}
\newcommand{\Ga}{\Gamma}
\newcommand{\epla}{\epsilon^\lambda_{0\tau}}
\newcommand{\eplab}{\epsilon^\lambda_{0\bar{\tau}}}
\newcommand{\eplax}{\epsilon^\lambda_{01}}
\newcommand{\eplay}{\epsilon^\lambda_{02}}

\newcommand{\bmu}{{\bm\mu}}

\newcommand{\be}{\begin{equation}}
\newcommand{\ee}{\end{equation}}

\newcommand{\bea}{\begin{eqnarray}}
\newcommand{\eea}{\end{eqnarray}}
\newcommand{\bd}{\begin{displaymath}}
\newcommand{\ed}{\end{displaymath}}
\newcommand{\ba}{\begin{array}}
\newcommand{\ea}{\end{array}}
\newcommand{\bi}{\begin{itemize}}
\newcommand{\ei}{\end{itemize}}
\newcommand{\bc}{\begin{center}}
\newcommand{\ec}{\end{center}}
\newcommand{\bfl}{\begin{flushleft}}
\newcommand{\efl}{\end{flushleft}}
\newcommand{\bfr}{\begin{flushright}}
\newcommand{\efr}{\end{flushright}}
\newcommand{\non}{\nonumber}

\newcommand{\bl}{\begin{aligned}}
\newcommand{\el}{\end{aligned}}

\newcommand{\hh}{\hat{h}}

\newcommand{\hZ}{\hat{Z}}

\newcommand{\hchi}{\hat{\chi}}

\newcommand{\bJ}{\bar{J}}

\newcommand{\bav}{\bar{V}}
\newcommand{\bartau}{\bar{\tau}}

\newcommand{\tiZ}{\tilde{Z}}

\newcommand{\tbt}{\tilde{\beta}}

\newcommand{\fs}{\frac{1}{2}}

\newcommand{\om}{i\omega_n}

\newcommand{\nul}{i\nu_l}

\newcommand{\ra}{\rangle}

\newcommand{\YB}{YbB$_{12}$}

\def\ket#1{\left\vert #1 \right\rangle}
\def\dg{^{\dagger}}

\def\bk{{\bf k}}  \def\bq{{\bf q}}

\def\bQ{{\bf Q}}  \def\bj{{\bf j}} 
  
 \def\bd{{\bf d}} \def\bS{{\bf S}} \def\bJ{{\bf J}}
 \def\bS{{\bf S}} 
\def\hbx{\hat{{\bf x}}}  

\def\bs{{\bf s}} \def\bJ{{\bf J}}

\def\dg{\dagger}

\def\no{\nonumber \\}

\def\ket{\rangle}
\def\={\!\!\!&=&\!\!\!}
\def\+{\!\!\!&&\!\!\!+~}
\def\-{\!\!\!&&\!\!\!-~}

\usepackage{color}
\usepackage[dvipsnames]{xcolor}
\usepackage{xcolor}
\usepackage[colorlinks=true,citecolor=blue]{hyperref}

\usepackage{hyperref,cleveref}

\begin{document}

\title{Optical  and magnetic excitations in the underscreened quasi-quartet Kondo lattice}

\author{Alireza Akbari$^{1,2,3}$}
\email{akbari@postech.ac.kr}
\author{Peter Thalmeier$^{2}$}
\affiliation{$^1$Max Planck POSTECH Center for Complex Phase Materials, POSTECH, Pohang 790-784, Korea}
\affiliation{$^{2}$Max Planck Institute for the  Chemical Physics of Solids, D-01187 Dresden, Germany}
\affiliation{$^3$Department of Physics, POSTECH, Pohang, Gyeongbuk 790-784, Korea}
\date{\today}

\begin{abstract}
The underscreened Kondo lattice consisting of a single twofold degenerate conduction band 
and a CEF split 4f- electron quasi-quartet has non-conventional quasiparticle dispersions obtained from  the constrained mean-field theory. An additional genuinely heavy band is found in the main hybridization band gap of the upper and lower hybridzed bands whose heavy effective mass is controled by the CEF splitting. Its presence should profoundly influence the dynamical optical and magnetic  response functions. In the former the onset of the optical conductivity is not the main hybridisation energy but the much lower Kondo energy scale which appears in the direct transitions to the additional heavy band. The dynamical magnetic response is also strongly modified by the in-gap heavy band which can lead to unconventional resonant excitations that may be interpreted as coherent CEF-Kondo lattice magnetic exciton bands. Their instability at low temperature signifies the onset of induced excitonic magnetism in the underscreened Kondo lattice.
\end{abstract}


\maketitle

\section{Introduction}
\label{sec:introduction}

Fundamental electronic properties of correlated f-electron compounds can be qualitatively understood within the 
Anderson lattice or Kondo-lattice (KL) models~\cite{newns:87,hewson:93,coleman:15,Otsuki:2009aa,Eder:2019aa}. In the strongly correlated limit (forbidden double occupancies) with 
large f-electron repulsion $U_{ff}\rightarrow\infty$ the slave-boson mean-field treatment with correlations simulated by a charge constraint on fermion and boson fields provides the most direct access to a description of renormalized quasiparticle bands. 
The combined effect of conduction (s-) and f- electron hybridzation as well as the f-electron correlation leads to two fundamental properties. Firstly the bands appear in (degenerate) pairs with a hybridazation gap existing between them for general \bk- points in the Brillouin zone (BZ). The size of the (indirect) effective gap is reduced to the order of the single ion Kondo temperature $T^*$. Secondly, due to this small energy scale in the range of a few meV  the quasiparticle bands close to the gap are very flat corresponding to large enhancement of the effective quasiparticle mass. 

The latter explains the thermodynamic and also transport properties of heavy fermion metals at low temperatures. In these materials, mostly Ce-intermetallics like, e.g. CeAl$_2$, CeB$_6$, CeCoIn$_5$ and many others the chemical potential is located in the flat part of the lower quasiparticle band. Due to residual quasiparticle interaction heavy fermion metals are prone to instabilities resulting, as in the above compounds, in exotic low temperature magnetic, multipolar~\cite{Shiina:1997aa,Portnichenko:2020aa} and superconducting phase transitions~\cite{thalmeier:05}.
In rare cases like the much discussed SmB$_6$ or YbB$_{12}$  borides~\cite{thalmeier:19} the chemical potential resides inside the hybridization gap leading to
 a Kondo insulator or semiconductor state (in the former due to the mixed valence~\cite{utsumi:17} of $\approx 2.5+$, it should be better termed mixed valence or hybridization gap insulator).
Likewise the low energy charge and spin response as represented by optical conductivity and inelastic neutron scattering (INS) can be qualitatively understood within the mean field slave boson approach of the Kondo lattices~\cite{riseborough:92,ikeda:96,riseborough:03,saso:06}, selfconsistent perturbation theory \cite{mutou:94} and also dynamical mean field technique~\cite{vidhy:03,chen:14}. In particular the appearance of a collective spin exciton resonance  observed in many f-electron materials (possibly superconducting or with hidden order) inside the hybridization gaps or those opened by symmetry breaking may be  interpreted within this approach~\cite{akbari:09,akbari:12,akbari:15,thalmeier:16}. 

Generally for these purposes the simplified SU(N) Kondo lattice model is employed. It assumes that the degeneracy N of localized (4f- or 5f-) states is the same as that of conduction electron states. Without crystalline electric field (CEF) effect the former is $(2J+1)$ and this may be quite large (J is the f-electron total angular momentum). In practice the CEF splitting reduces the f-electron degeneracy  to N=2, 4 (the latter only in cubic environment). However, if the splitting of CEF ground and first excited states is moderate both are involved in the Kondo screening leading to the heavy quasiparticle bands, thus possibly invoking a larger quasi-degeneracy. This poses a problem for the straightforward application of the SU(N) KL model. For general wave vector in the Brillouin zone  conduction electron states are at most twofold Kramers degenerate when time reversal symmetry holds and their spin orbit coupling is neglected. Higher degeneracy can only appear at symmetry positions. Therefore for $N>2$ the genuine KL model is rather artificial since degeneracies of f-and conduction states no longer matches. This problem can be treated for the impurity model~\cite{nozieres:80,coleman:15}  but in the lattice it is rather difficult to analyze properly. 

Physically in most cases it is more reasonable to assume an `underscreened' model with higher f-electron than conduction electron degeneracy, where the former may actually be of pseudo-type, i.e. with a small CEF splitting of the same order as the Kondo temperature. Such a model has recently been investigated in detail in view of its quasiparticle spectrum and how the latter deviates from the canonical two N-fold degenerate hybridized bands  of the genuine SU(N) KL model (see Ref.\onlinecite{thalmeier:18} and earlier work in Refs.~\onlinecite{perkins:07,thomas:11,thomas:14}). Specifically a quasi-quartet  KL model for f-states was studied which is e.g. relevant for the (reduced) $\Gamma_6-\Gamma_7$  CEF- level scheme in YbRu$_2$Ge$_2$ and similar tetragonal compounds~\cite{jeevan:06,takimoto:08,jeevan:11}, hybridizing with a simple twofold degenerate conduction band. As an important  result it was obtained that in the underscreened case an additional heavy quasiparticle band appears within the main hybridization gap whose dispersion is controled by the interplay of CEF splitting and Kondo screening. This gives the model a much richer low energy band structure than the common SU(N) model with several more discernible hybridization gaps.
It would be highly desirable to probe this unconventional KL quasiparticle spectrum with inelastic low energy probes such as optical conductivity, inelastic neutron scattering (INS) and STM-techniques, since ARPES does not have the resolution to probe such 
subtle features in heavy bands. Actually STM-QPI technique has given indications in two heavy fermion metals that the hybridized band structures are more complex than suggested by the common SU(N) KL model with upper and lower hybridized branches~\cite{allan:13,Giannakis:2019aa,schmidt:10}.\\ 

It is the main purpose of the present work to study in detail inelastic low energy response of the underscreened quasi-quartet KL model based on the results of the previous work~\cite{thalmeier:18} to predict the signatures of the additional heavy band of this model in optical conductivity $\sigma(\omega)$  and inelastic neutron scattering $S(\bq,\omega)$. We show that their signatures appear as additional shoulders and peaks in the frequency dependencies of these experimental quantities and for INS have a distinctly dispersive behaviour. We argue that our results on optical conductivity suggest a simple explanation for unconventional behaviour of this quantity found in the Kondo insulator YbB$_{12}$ \cite{okamura:05}. Furthermore we discuss the possibility of a hybrid CEF-Kondo magnetic exciton mode in the dynamical magnetic response in the  heavy bands of the underscreened KL model. We also give a qualitative discussion for the appearance of induced excitonic magnetism due to  dominating non-diagonal exchange in the quasi-quartet and how the corresponding instability criterion is influenced by the Kondo screening for CEF split f-electrons.


\section{Quasiquartet KL model}
\label{sec:model}

\begin{figure}
\includegraphics[width=0.99\columnwidth]{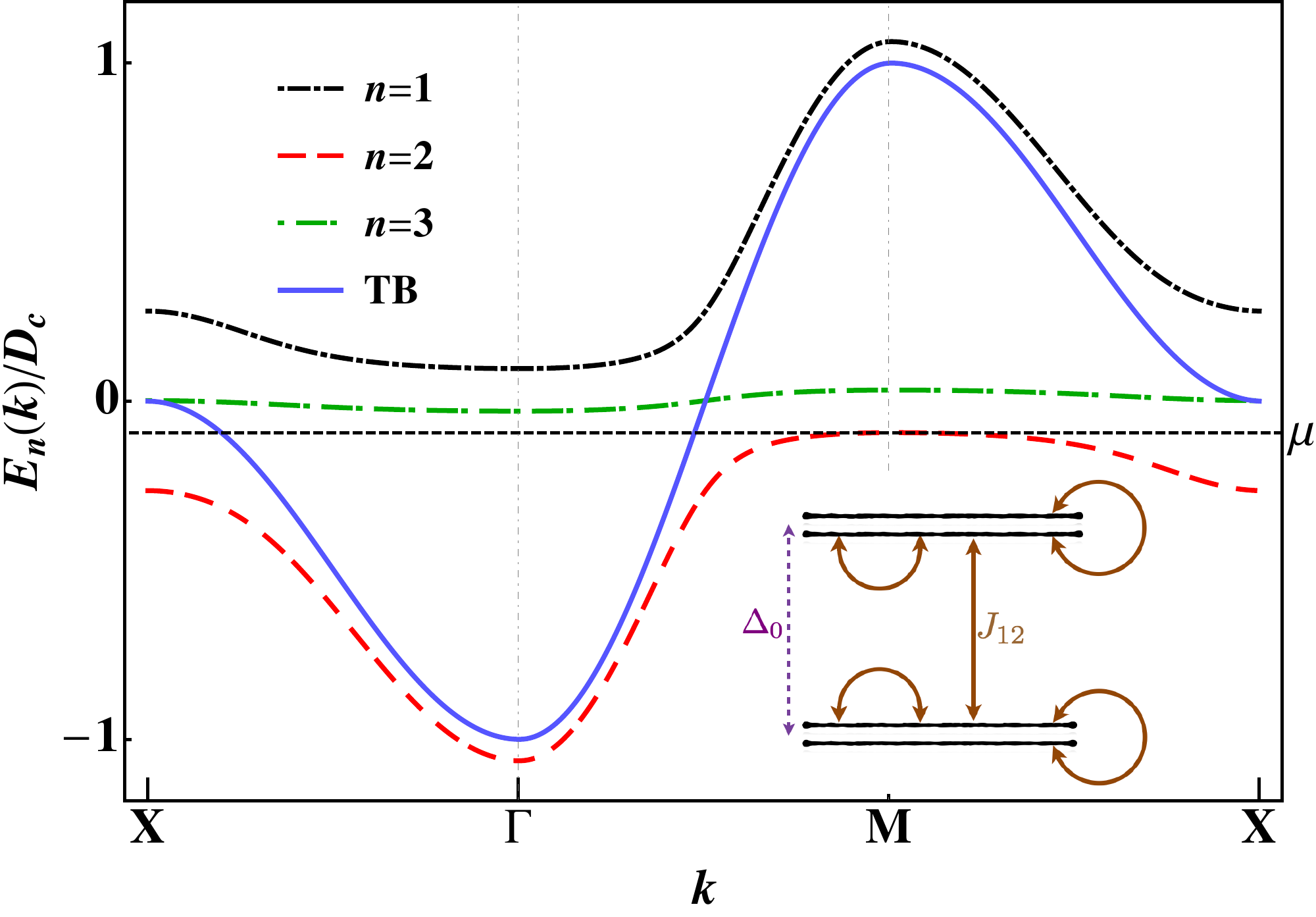} 
\caption{Quasiparticle dispersion unfolded in the 2D BZ with $\Gamma(0,0)$, $X(\pi,0)$ and $M(\pi,\pi)$.
The lower right inset shows the quasiquartet model (orbital index $\tau =1,2$) denotes ground and excited Kramers doublets rexpectively). Its effective Kondo couplings to conduction electrons are given by
$J^\perp_1=c_{11}J_0; \; J^\perp_2=c_{22}J_0; \; J_{12}=\frac{1}{\sqrt{2}}c_{12}J_0$ and 
$J^z_1=c^z_{11}J_0; \; J^z_2=c^z_{22}J_0; \; J^z_{12}=0$ where $J_0=(g_J-1)I_{ex}$ is the bare Kondo exchange in Eq.~(\ref{eqn:cfHam}). Here $\mu=-0.094$.}
\label{fig:BZdisp}
\end{figure}

The quasiquartet model is illustrated in the inset of Fig.~\ref{fig:BZdisp} showing the two CEF-split Kramers doublets
$(\tau=1,2)$ which interact with the conduction electrons that are scattered both elastically $(\tau\leftrightarrow\tau)$  and inelastically $(1\leftrightarrow 2)$. The basic Hamiltonian of the lattice of quasi quartets interacting with the single (doubly Kramers degenerate) conduction band is then
 \be
 {\cal H}_{\rm K}= {\cal H}_{\rm CEF}+ \sum_{\bk\si}\epsilon_\bk c^\dag_{\bk\si}c_{\bk\si}+(g_J-1)I_{ex}\sum_i\bs_i\cdot\bJ_i ,
 \label{eqn:cfHam}
 \ee
where $\epsilon_\bk=-(D_c/2)(\cos k_x +\cos k_y)$ is the conduction band dispersion with band width $2D_c$, $g_J$ the f-electron g-factor corresponding to the CEF split total f-electron angular momentum $(\bJ)$ multiplet and $I_{ex}$ the exchange coupling strength. 
The (isotropic) exchange part (which is of rank 1 in $J^\mu_i, (\mu=x,y,z)$ is one term extracted from a Schrieffer-Wolff transformation of the original Anderson model. There are additional terms obtained that couple conduction electrons to more general localized 4f-operators \cite{kusunose:16} such as quadrupoles (rank 2) and octupoles (rank 3) that are supported in the quasi-quartet CEF system \cite{takimoto:08}. Due to momentum dependent form factors these terms correspond to anisotropic conduction electron scattering which complicate the treatment of selfconsistency equations in the slave-boson treatment. Therefore we restrict to the above simple isotropic exchange term.

The Kondo lattice model of Eq.~(\ref{eqn:cfHam}) is of the underscreened type because there are only $N=2$ conduction states that interact with $2N=4$ localized quasi-quartet f-states. The heavy quasiparticle spectrum of this model was studied in detail in Ref.~\onlinecite{thalmeier:18} using a fermionic representation of $ {\cal H}_{\rm K}$ and treating it within a constrained mean field theory. Using the spinors  $\Psi^\dag_{\bk\si}=(c^\dag_{\bk\si},f^\dag_{1\bk\si},f^\dag_{2\bk\si})$ where $c^\dag_{\bk\si}$ and $f^\dag_{\tau\bk\si}$ create conduction and f-electrons this leads to a bilinear fermionic mean field Hamiltonian 
\be
\tilde{\cal H}_{mf}^\lam=\sum_{\bk m}\Psi^\dg_{\bk m} \hh_\bk \Psi_{\bk m}; \;\;\;
\hh_{\bk}=
\left(
 \begin{array}{ccc}
\epsilon_\bk& \bav_1 & \bav_2 \\
\bav_1& \epsilon_{01}^\lam& 0 \\
\bav_2& 0& \epsilon_{02}^\lam
\end{array}
\right).
\label{eqn:bilHam}
\ee
Here $\lambda$ is the effective f-level and $\bav_\tau$ its effective hybridization with conduction electrons  in the fermionic representation.  These quantities are determined by f- occupation constraint $n_f=1$, conduction electron number $n_c$ and a selfconsistency relation~\cite{thalmeier:18}. We define $\De=\fs\De_0$ so that the CEF split effective f-level energies are $\epsilon^\lam_{01}=\lam-\De;\;\; \epsilon^\lam_{02}=\lam+\De$.  Furthermore $\epsilon^\lam_\bk =\epsilon_\bk-\lam$ will be used. The diagonalization of the mean field Hamiltionian leads to three quasiparticle bands
\be
\bl
E_{1\bk}&=\lam +\frac{1}{3}\epsilon^\lam_\bk+2\sqrt[3]{r_\bk}\cos
(\frac{\phi_\bk}{3}),
\\
E_{2\bk}&=\lam +\frac{1}{3}\epsilon^\lam_\bk+2\sqrt[3]{r_\bk}\cos(\frac{\phi_\bk}{3}+\frac{2\pi}{3}),
\\
E_{3\bk}&=\lam +\frac{1}{3}\epsilon^\lam_\bk+2\sqrt[3]{r_\bk}\cos(\frac{\phi_\bk}{3}+\frac{4\pi}{3}).
\label{eqn:qpbands}
\el
\ee
where we used the auxiliary quantities 
\be
\bl
r_\bk
&
=\bigl(\frac{1}{3}[(\De^2+\bav^2)+\frac{1}{3}\epsilon^{\lam 2}_\bk]\bigr)^\frac{3}{2},
\\
\cos\phi_\bk
&=\frac{1}{2r_\bk}
\Bigl[
\frac{1}{3}\epsilon^\lam_\bk[\frac{2}{9}\epsilon^{\lam 2}_\bk +(\De^2+\bav^2)]
-\De(\epsilon^\lam_\bk\De+\delta_s)
\Bigr],
\label{eqn:auxiliary}
\el
\ee
with the definition $\bav^2=\bav^2_1+\bav^2_2$  and  $\delta_s=\bav^2_1-\bav^2_2$. The essential problem is the
selfconsistent determination of the effective  f-level position $\lambda$ with respect of Fermi energy $\mu$  and the effective
hybridization $\bav_\tau$ of CEF split f-states~\cite{thalmeier:18}. An example for the structure of quasiparticle bands in the symmetric case $\bav_1=\bav_2$ is shown in Fig.~\ref{fig:BZdisp}. There are two hybridized bands $E_{1,2\bk}$ of partly conduction and f- electron character that changes when crossing the Brillouin zone (BZ). They have alternating flat portions corresponding to heavy 
effective mass. On the other hand the central overall flat band  $E_{3\bk}$ has mostly f- electron character with only small c-electron admixture that causes the small band width given by
\bea
W_3=(T^{*2}+\De_0^2)^\fs-T^*,
\eea
which is nonzero only for  finite CEF splitting $\De_0$. Here, in the symmetric case
\bea
T^*(\De)=\frac{\bav^2}{D_c}\simeq (\lam-\mu)-\De
\label{eqn:TKondo}
\eea
is the low energy (Kondo) scale for the heavy bands that determines their mass and hybridization gaps. It is obtained from solving the selfconsistency equation for $\lambda-\mu$. The former are given by 
$\frac{m_{1,2}^*}{m_b}=\frac{D_c}{T^*}$ in the heavy mass part of the BZ and $\frac{m^*_3}{m_b}=\frac{T^*D_c}{\De^2}$ throughout the BZ so that $\frac{m^*_3}{m^*_{1,2}}=\bigl(\frac{T^*}{\De}\bigr)^2$ (for $\Delta\rightarrow 0$ the central $E_{3\bk}$ band becomes flat and $m^*_3$ diverges). Here $m_b=\frac{\hbar^2k_F}{D_c}$ is the
unrenormalized conduction band mass. Furthermore the main {\it indirect} hybridization gap is given by $\De_{h1}^{in}=E_{10}-E_{2Q}\simeq T^*+(T^{*2}+\De_0^2)^\fs$. The whole heavy $E_{3\bk}$ band lies within this gap. There are additional indirect and direct hybridization gaps to be identified as discussed in detail in Ref.\onlinecite{thalmeier:18}. These features are nicely illustrated
by the example of heavy band structure in Fig.~\ref{fig:BZdisp}(b) for the (particle-hole) symmetric case. We emphasize that these
bands should not be considered like ordinary non-interacting bands that can be rigidly filled up the chemical potential $\mu$ with an
arbitrary position. This is not true due to the constraint $n_f=1$ enforced by the strong on-site correlations $(U_{ff}\rightarrow\infty$). Therefore when the chemical potential changes the effective level $\lambda$ and hence the hybridization gap structure in Fig.~\ref{fig:BZdisp} is tied and moving along with the chemical potential. This means the chemical potential always has to be in close vicinity to the hybridization gap structure in the DOS~\cite{thalmeier:18}.

%
%

\section{Optical conductivity}
\label{sec:opticalcond}

The optical conductivity in Kondo lattice compounds exhibits two distinct features \cite{coleman:15}.
In the metallic case when the chemical potential is pinned in the flat part of the lower band ($n=2$ in Fig.~\ref{fig:BZdisp})
a quasielastic Drude peak appears whose weight is $\sim(m_b/m^*_2)=T^*/D_c$ is suppressed due to the mass renormalization and the width
is determined by the phenomenological quasiparticle relaxation rate in that band. This is the conventional Fermi-liquid type part.
The more interesting high frequency part connected with the detailed hybridisation gap structure and the inelastic optical transitions across those gaps is not described by this phenomenological approach. It requires the full microscopic theory based on the (underscreened) Kondo lattice model which is developed in this section.
The microscopic expression for the optical conductivity (real part) is derived from the response  function associated with the
$(\bq=0)$ conduction electron current 
\be
\bold{j}=e\sum_{\bk\si}\nabla_\bk\epsilon_\bk c^\dag_{\bk\si}c_{\bk\si}.
\ee
Then the conductivity $(\bj\parallel\hbx)$ is obtained as ~\cite{coleman:15,lee:15}
\be
\bl
&
\si(\omega)=\frac{1}{\omega}
{\rm Im}
\Big[
\Pi(\nul)
\Big]_{\nul=\omega+i 0^+},
\label{eqn:sigma}
\el
\ee
with
\be\no
\bl
\Pi(\nul)=-T\sum_{\bk,\om}(v^{x}_{\bk})^2G_c(\bk,\om)G_c(\bk,\om+\nul),
\el
\ee
where $v^{x}_{\bk}=\nabla_{k_x}\epsilon_\bk$ is the group velocity. Using the spectral representation
\be
G_{c}(\bk,\om)=\int_{-\infty}^\infty d\omega\frac{\rho_{c}(\bk,\omega)}{\om-\omega},
\label{eqn:cspectral}
\ee
where $\rho_{c}(\bk,\omega)$ is the renormalized conduction electron spectral function given by
\be
\rho_c(\bk,\omega)=\delta(\omega-\Sigma_c(\omega)-\epsilon_\bk);\;\;\;
\Sigma_c(\om)=\Sigma_{\tau}\frac{\bav^2_\tau}{\om-\epla},
\label{eqn:conspec}
\ee
Here $\Sigma_c(\om)$ denotes the conduction electron self energy due to hybridization with the two f-orbitals $(\tau=1,2)$.
Its evaluation leads to a sum of delta functions at the quasiparticle energies $(\beta=1-3)$ weighted by c-electron residua:
\be
\bl
&\rho_{c}(\bk,\omega)
=\sum_\bt\tiZ_{\bt\bk}\delta(\omega-E_{\bt\bk});
\\
&
\tiZ_{\bt\bk}
=\frac{|E_{\bt\bk}-\eplax| |E_{\bt\bk}-\eplay|}{\Pi_{\al\neq\bt}|E_{\bt\bk}-E_{\al\bk}|},
\label{eqn:cresidua}
\el
\ee
The explicit forms of the  $\tiZ_{\bt\bk}$ is given in Appendix \ref{sec:app1}.
Inserting  Eq.~(\ref{eqn:cspectral}) into Eq.~(\ref{eqn:sigma}) and averaging out the velocity we obtain
\be
\bl
&\Pi(\nul)=
\\
&
\frac{\omega_{pl}^2}{4\pi}
\sum_\bk\int d\omega' d\omega''
\rho_{c}(\bk,\omega')\rho_{c}(\bk,\omega'')
\frac{f(\omega')-f(\omega'')}{\nul+\omega''-\omega'},
\;\;\;
\label{eqn:polar}
\el
\ee
where $\omega_{pl}^2=4\pi n_ce^2/m_b$ is the plasma frequency with $n_c$ and, $m_b$ the conduction electron density
and effective band mass, respectively. As for the magnetic susceptibilities (Sec.~\ref{sec:sus0}) we may evaluate this
expression using the explicit conduction electron spectral function given above.
Then, using Eqs.~(\ref{eqn:sigma},\ref{eqn:polar}) and the residual weights given in Eq.~(\ref{eqn:tilres}) we finally obtain for the optical conductivity
\be
\bl
\si(\omega)=&\frac{\omega_{pl}^2}{4\omega} \sum_{\bt\tbt}\sum_\bk 
\\&\hspace{-0.5cm}
\Big[
\tiZ_{\bt\bk}\tiZ_{\tbt\bk}
\Big(f(E_{\tbt\bk})-f(E_{\bt\bk})
\Big)
\delta
\Big(\omega-(E_{\bt\bk}-E_{\tbt\bk})
\Big)
\Big],
\;\;\;
\label{eqn:siginter}
\el
\ee
%
Because the chemical potential is closely located below the upper edge of the lowest band due to the 
constraint $n_f =1$~\cite{thalmeier:18} there wil be a Drude term from the corresponding intra-band transitions.  In the limit $T\rightarrow 0$ $f(E_{\tbt\bk})=\Theta_H(E_{\tbt\bk})$ the inter-band contributions  $(\bt\neq\tbt)$ in Eq.~(\ref{eqn:siginter}) contain only pairs $(\bt\tbt) = (2,3)$, and $(2,1)$ since only band $\bt=2$ is occupied and $\tbt=3,1$ are empty.
Thus the optical conductivity will have several threshold frequencies given by the various hybridization gaps in Fig.~\ref{fig:BZdisp}, all of them corresponding to {\it direct} transitions (\bq=0). 
Because the quasipartcle $(E_{\beta\bk})$ and residual $(\tiZ_{\beta\bk})$  \bk~- dependencies in Eqs.~(\ref{eqn:qpbands},\ref{eqn:cresidua}) stem entirely from the conduction electron dispersion we may convert the \bk-summation in Eq.~(\ref{eqn:siginter}) into an integral over the 
bare conduction electron DOS according to
\be
\bl
\si(\omega)=&
\frac{\omega_{pl}^2}{4\omega}\sum_{\bt\tbt=23,21}\int_{-D_c}^{Dc}d\epsilon
\\
&
\Big[
\rho_c^0(\epsilon)
\tiZ_{\bt}(\epsilon)\tiZ_{\tbt}(\epsilon)
\delta
\Big(
\omega-
[E_{\bt}(\epsilon)-E_{\tbt}(\epsilon)]
\Big)
\Big].
\el
\ee
As models the square and tight-binding DOS have been used before~\cite{thalmeier:18} corresponding to a  band width $2D_c$. In the former case the DOS may be taken outside the integral as the constant $\rho_c^0=1/2D_c$.

\begin{figure}
\includegraphics[width=0.95\columnwidth]{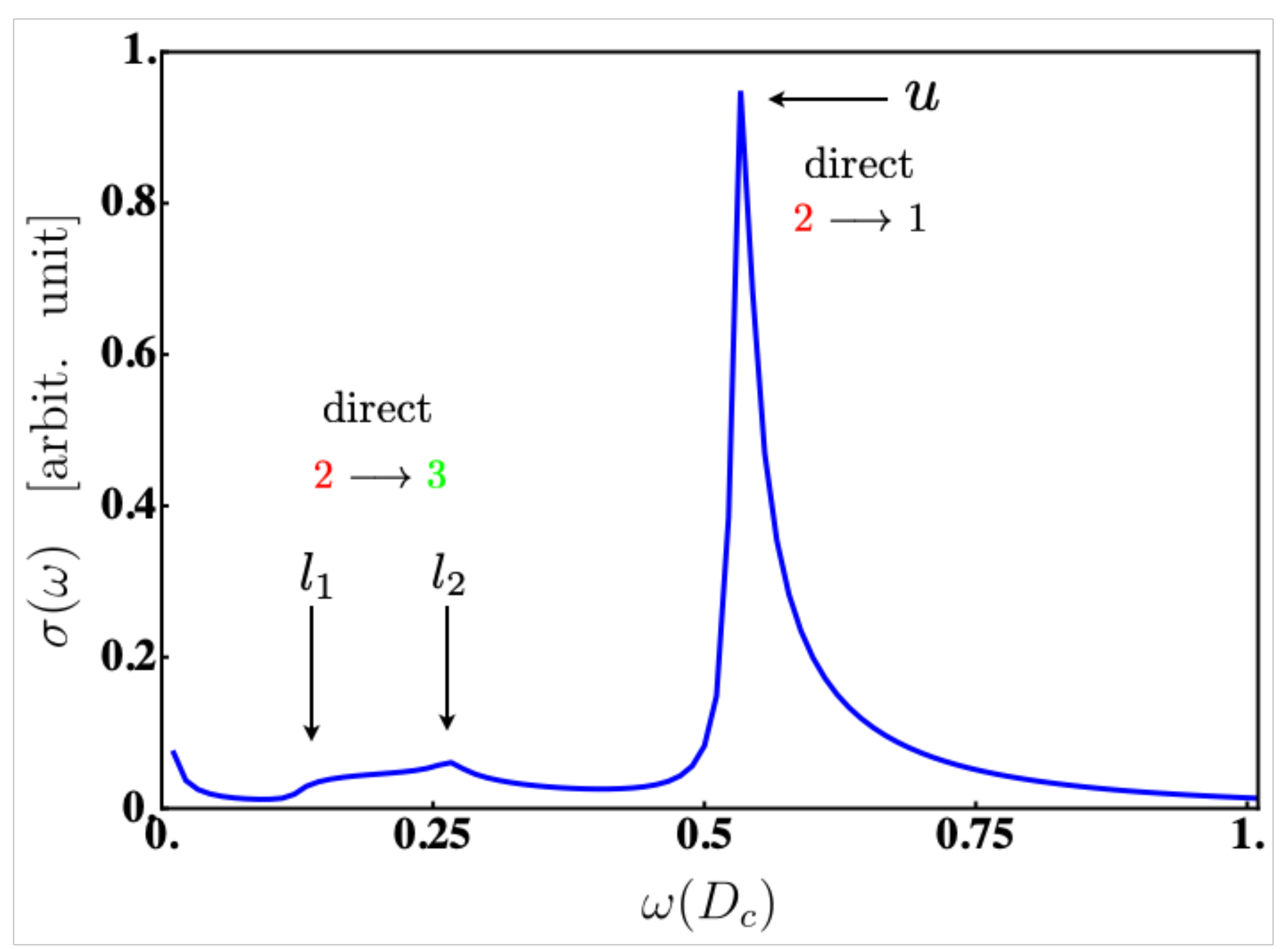}
\vspace{-0.25cm}
\caption{
Frequency dependence of the optical conductivity, $\sigma(\omega)$. It shows three features: i) A $\omega\approx 0$ Drude peak due to chemical potential $\mu\leq E_{2\bQ}$ $(\bQ=(\pi,\pi)$; M-point) close to upper edge of $E_{2\bk}$. ii) A small low energy inelastic peak between $\omega_{l_1}<\omega<\omega_{l_2}$ due to direct transitions at M($l_1$) (small hybridzation gap $\De_{h3}^{d}$) and X($l_2$) between $n=2,3$ bands. iii) a large high energy inelastic peak at $\omega_u$ due to direct transition at X and center of $\Gamma$M corresponding to large hybridization gap  $\De_{h1}^d=2(\bav^2+\De^2)^\fs$ between $n=2,1$ bands. }
\label{fig:optcond}
\end{figure}

For the numerical calculations we use directly the general expression Eq.~(\ref{eqn:siginter}). 
The results for $\sigma(\omega)$  are presented in Fig.~\ref{fig:optcond}. There is a small Drude peak from the lowest band whose width is determined by the small imaginary part $(\gamma =0.001D_c$) used in the integration. Most importantly {\it two} inelastic peaks in the frequency dependence are visible, corresponding to the {\it two direct} $(\bq=0)$ hybridization gaps identified in the quasiparticle spectrum~\cite{thalmeier:18} and visble in Fig.~\ref{fig:BZdisp}. In the terminology of Ref.~\onlinecite{thalmeier:18} the lower one starting at $\omega_{l1}\sim 0.13D_c$ corresponds to 
$\De_{h3}^{d}=E_{3\bQ}-E_{2\bQ}=D_c\bigl[(\frac{\bav^2}{D_c^2})^2+\frac{\De_0^2}{D^2_c}\bigr]^\fs =\bigl(T^{*2}+\De^2_0\bigr)^\fs \equiv E_{10}-E_{30}=\De_{h2}^d$,
while the upper one at  $\omega_u\sim 0.5D_c$ stems from $\De_{h1}^d=E_{1\bQ'}-E_{2\bQ'}=2(\bav^2+\De^2)^\fs$.
The former is determined by the low energy Kondo and CEF-splitting energy scales $T^*=\frac{\bav^2}{D_c}$ and $\Delta$, respectively while the upper one by the larger effective hybridisation scale $2\bav$ since $\Delta\ll\bav$. The lower peak is much less pronounced because it is associated with optical transitions from the occupied to the central heavy band whose Bloch functions have only small c-electron content \cite{thalmeier:18}.
The presence of two energy scales and peaks in the optical conductivity due to the posssibility of two direct transition
is a decisive difference to the conventional Kondo lattice model which exhibits only the `high' energy scale of  $\De_{h1}^d\simeq2\bav$  in $\sigma(\omega)$. The Kondo scale $T^*$ is not associated with any direct feature in $\sigma(\omega)$ because in the fully screened SU(N) KL it only appears for the {\it indirect} transitions with BZ-boundary momentum transfer \bQ. There is indeed some evidence that the two energy scales of the more realistic underscreened model may have been observed experimentally (Sec.~\ref{sect:discussion}).

%
%

\section{Bare magnetic susceptibilities}
\label{sec:sus0}

For the calculation of the dynamic magnetic response we first need to calculate the bare physical magnetic susceptibilities
coming from particle hole excitations in the heavy bands due to the dynamics of magnetic moments $g_J\mu_B\bJ$. These are combinations of the pseudospin susceptibilities. In terms of the pseudospin operators for the two Kramers doublets ($\tau,\tau'=1,2$)~\cite{takimoto:08}
\bea
S^\alpha_{\tau\tau'}=\fs\sum_{\si\si'}f^\dag_{\tau\si}\hat{\si}^\alpha_{\si\si'}f_{\tau'\si'}^{},
\label{eqn:pseudo}
\eea
the total angular momentum operators, constricted to the quasiquartet CEF system may be written as 
\be
\bl
J^z=&c^z_{11}S^z_{11}+c^z_{22}S^z_{22},
\\
J^\pm=&c_{11}S^\pm_{11}+c_{22}S^\pm_{22}+c_{12}\frac{1}{\sqrt{2}}(S^\pm_{12}+S^\pm_{21}).
\label{eqn:JSequiv}
\el
\ee
Here the coefficients $c^z_{\tau\tau'}$ and $c_{\tau\tau'}$ are determined by the parameters of the
CEF potential or their eigenstates~\cite{takimoto:08,thalmeier:18}  (see Appendix \ref{sec:app2} and Table~\ref{table:1} for more details).\\

The in-plane $(\mu =x,y\; \text{or}\; \perp)$ and out of plane $(\mu =z\; \text{or} \parallel)$ suszeptibilities of the physical moment operators $\bmu=g_J\mu_B\bJ$ are related to
their reduced expressions according to 
\bea
\chi^\mu(\bq,\nul)=g_J^2\mu_B^2\hchi^\mu(\bq,\nul).
\label{eqn:sus}
\eea
Here $\mu=x,y (\perp)$ cartesian components are equivalent due to tetragonal symmetry. Using Eq.~(\ref{eqn:JSequiv}) the two independent components $(\mu=\perp,z)$ of the reduced susceptibility $\hchi^l(\bq,\nul)$ may then be expressed by the pseudospin susceptibilities according to 
\be
\bl
&
\hchi^\parallel(\bq,\nul)
=\sum_\tau (c^z_{\tau\tau})^2\hchi_{\tau\tau}(\bq,\nul);
\\
&\hchi^\perp(\bq,\nul)
=
\\&
\sum_\tau
\Big[
 (c_{\tau\tau})^2\hchi_{\tau\tau}(\bq,\nul)
 \!+\!
\fs(c_{12})^2
[\hchi^a_{\tau\bartau}(\bq,\nul)+\hchi^b(\bq,\nul)]
\Big].
\label{eqn:susphys}
\el
\ee
The second term in the last equation $\sim c_{12}^2$ is the inter-orbital or vanVleck contribution due to nondiagonal matrix elements between the two CEF doublets (inset in Fig.~\ref{fig:BZdisp}).
In the bubble approximation~\cite{coleman:15} the bare pseudospin susceptibilities may then be expressed via the 
Green's functions of fermionic variables in the pseudospin representation of Eq.~(\ref{eqn:pseudo}) according to
\be
\bl
\hchi_{\tau\tau}(\bq,\nul)
&=-\fs T\sum_{\bk,\om}G_{f\tau}(\bq',\om+\nul)G_{f\tau}(\bk,\om);
\\
\hchi^a_{\tau\bartau}(\bq,\nul)
&=-\fs T\sum_{\bk,\om}G_{f\tau}(\bq' ,\om+\nul)G_{f\bartau}(\bk,\om);
\\
\hchi^b(\bq,\nul)
&=-\fs T\sum_{\bk,\om}B(\bq' ,\om+\nul)B(\bk,\om),
\label{eqn:susred}
\el
\ee
where $\bq'=\bk+\bq $.
For further evaluation we use the spectral representation of Green's functions according to
\be
\bl
G_{f\tau}(\bk,\om)
&=\int_{-\infty}^\infty d\omega\frac{\rho_{f\tau}(\bk,\omega)}{\om-\omega};
\\
B(\bk,\om)
&=\int_{-\infty}^\infty d\omega\frac{\rho_B(\bk,\omega)}{\om-\omega},
\label{eqn:Greenspectral}
\el
\ee
where the f-electron spectral densities are given by~\cite{thalmeier:18}
\be
\bl
\rho_{f\tau}(\bk,\omega)
&=\frac{\bav^2_\tau}{(\omega-\epla)^2}\rho_c(\bk,\omega);
\\
\rho_B(\bk,\om)
&=\frac{\bav_1\bav_2}{(\om-\eplax)(\om-\eplay)}\rho_c(\bk,\omega). 
\label{eqn:spectral}
\el
\ee
Using Eq.~(\ref{eqn:cresidua}) they are evaluated as 
\be
\bl
\rho_{f\tau}(\bk,\omega)
&
=\sum_\bt\hZ^\tau_{\bt\bk}\delta(\omega-E_{\bt\bk});
\\
\rho_{B}(\bk,\omega)
&
=\sum_\bt\hZ^{B}_{\bt\bk}\delta(\omega-E_{\bt\bk}).
\label{eqn:specex}
\el
\ee
with the residual weights (see also Appendix \ref{sec:app1}) given by
\be
\bl
\hZ^\tau_{\beta\bk}
&=\frac{\bav^2_\tau |E_{\beta\bk}-\eplab|}{\Pi_{\al\neq\bt}|E_{\bt\bk}-E_{\al\bk}||E_{\bt\bk}-\epla |};
\\
\hZ^B_{\beta\bk}
&=\frac{\bav_1\bav_2\si_{\bt\bk}}{\Pi_{\al\neq\bt}|E_{\bt\bk}-E_{\al\bk}|},
\label{eqn:residua}
\el
\ee
and the definition of
\be
\si_{\bt\bk}={\rm sign}\Big[(E_{\bt\bk}-\eplax)(E_{\bt\bk}-\eplay)\Big]=\pm 1.
\label{sign}
\ee
Inserting Eq.~(\ref{eqn:Greenspectral}) into Eq.~(\ref{eqn:susred}) and using the explicit form of spectral weights  in Eqs.~(\ref{eqn:spectral}~and~\ref{eqn:conspec}) and their residual form  to carry out the frequency integrations we obtain:
\be
\bl
\hchi_{\tau\tau}(\bq,\nul)
=&\fs\sum_{\bt\tbt}\sum_\bk\hZ^\tau_{\bt\bq' }\hZ^\tau_{\tbt\bk}
\frac{f(E_{\bt\bq' })-f(E_{\tbt\bk})}{\nul+E_{\tbt\bk}-E_{\bt\bq' }};
\\
\hchi^a_{\tau\bartau}(\bq,\nul)
=&\fs\sum_{\bt\tbt}\sum_\bk\hZ^\tau_{\bt\bq' }\hZ^{\bartau}_{\tbt\bk}
\frac{f(E_{\bt\bq' })-f(E_{\tbt\bk})}{\nul+E_{\tbt\bk}-E_{\bt\bq' }};
\\
\hchi^b(\bq,\nul)
=&\fs\sum_{\bt\tbt}\sum_\bk\hZ^B_{\bt\bq' }\hZ^B_{\tbt\bk}
\frac{f(E_{\bt\bq' })-f(E_{\tbt\bk})}{\nul+E_{\tbt\bk}-E_{\bt\bq' }}.
\label{eqn:susres}
\el
\ee
Here the \bk - summation runs over the 2D BZ and the $\bt,\tbt =1-3$ summation over the three quasiparticle bands
of Eq.~(\ref{eqn:qpbands}) comprising in principle three intra-band and three inter-band transitions. However due to
the constraint~\cite{thalmeier:18} $n_f=1$ the chemical potential lies in the lowest band $E_{2\bk}$ and then only one
intraband $2\leftrightarrow 2$ and two interband transitions $2\leftrightarrow 1,3$ contribute to the susceptibilities.\\

%
\begin{figure}
\includegraphics[width=1\columnwidth]{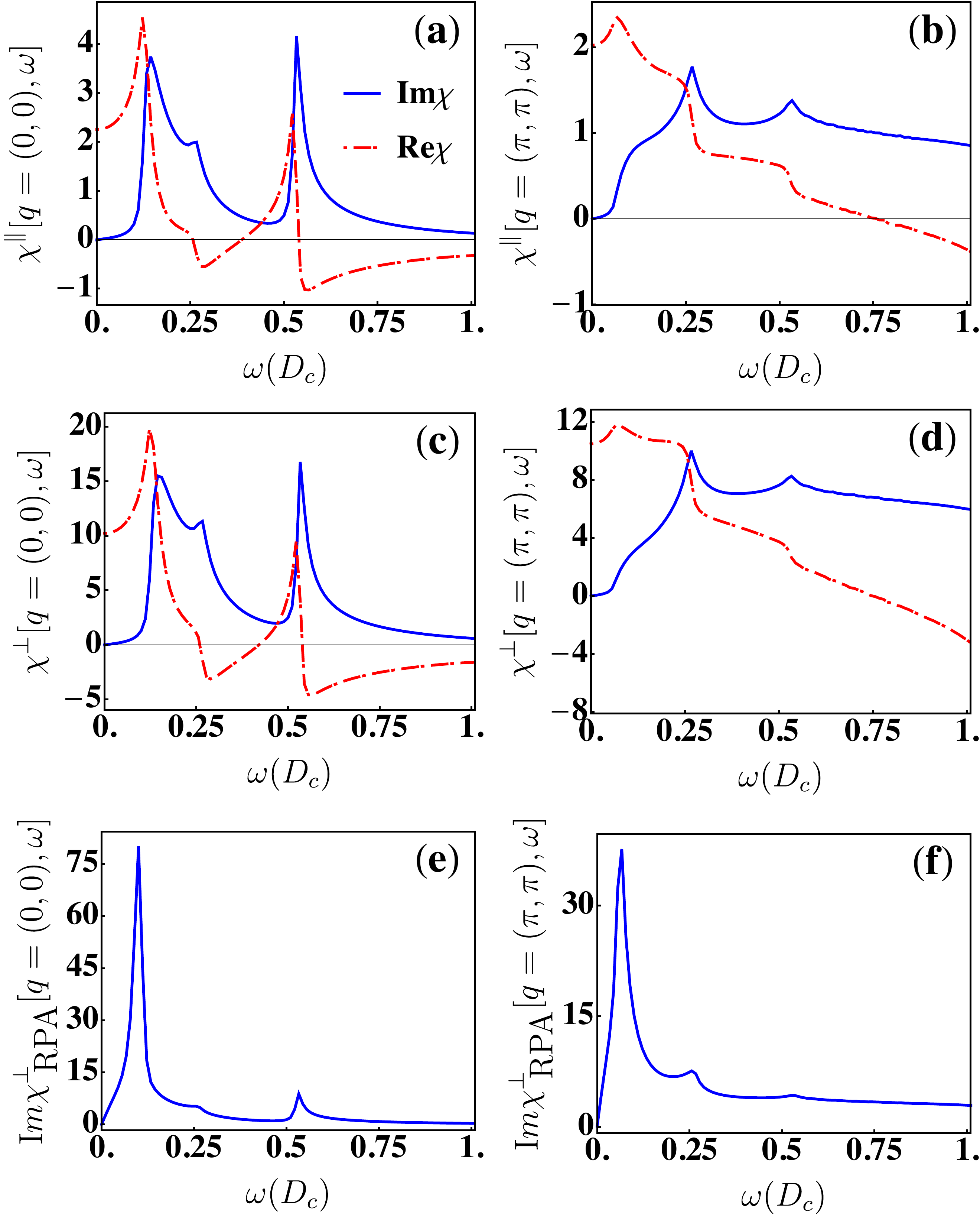}
\caption{
(a-d) Frequency dependence of bare  susceptibilities: (a,b) show the  in plane ($\parallel$)
 and (c,d) the  out of plane ($\perp$) magnetic response functions, 
 at $\bq=(0,0)$ and $\bq=(\pi,\pi)$, respectively.
Peak features at $\omega_{l}$ and $\omega_u$ for both $\bq=0,\bQ$ stemming from large and small hybridization gaps in Fig.~\ref{fig:BZdisp} can be discerned.
  (e,f) The frequency dependence of the perpendicular  RPA  susceptibilities, 
 at $\bq=(0,0)$ and $\bq=(\pi,\pi)$ for exchange function parameters $I^{\parallel,\perp}_0 =1/11.25=0.09, \Gamma=0.5$ [see Eq.~(\ref{eqn:intersite}))]. A collective exciton mode peak appears inside the lowest hybridization gap.
}
\label{fig:ChiPer}
\end{figure}
%

The results for the bare susceptibilities for both moment directions and for BZ center and boundary wave vectors are shown in Fig.~\ref{fig:ChiPer}(a,b) for $\parallel$ and Fig.~\ref{fig:ChiPer}(c,d) for $\perp$ where we plotted both real and imaginary parts.
For both moment directions the transitions across the two {\it direct} hybridsation gaps (Sec.~\ref{sec:opticalcond}) show up clearly as sharp separate peaks in the spectrum (imaginary part, blue)  associated with singular behaviour of the real part (red) at the zone center $\bq=0$. On the other hand for the zone boundary wave vector $\bQ=(\pi,\pi)$ a larger manyfold of {\it indirect} transitions is possible and the spectrum is more spread out in frequency, although certain individual idirect gap excitation energies are still discernible. This result leads one to consider identification of the two-peak structure not only in optical conductivity but also in inelastic neutron scattering that probes the magnetic response functions, albeit of the interacting system considered in the next section.

\section{Dipolar RPA dynamic susceptibility and spectrum}
\label{sect:RPAresponse}

From the two-impurity Kondo models it is known that it induces two competing effects: The on-site screening of moments that
tend to form a singlet ground state and creation of  effective inter-site (RKKY)-type couplings that prefer to align the moments to a magnetically ordered ground state in the lattice. The latter may be obtained in second order perturbation theory from Eq.~(\ref{eqn:cfHam}) by eliminating conduction electrons. In a similar way more generalised inter-site multipolar interactions are generated for the (quasi-)quartet system if one includes the higher rank($2,3)$- quadrupolar and octupolar terms in  Eq.~(\ref{eqn:cfHam}). They can favor hidden order ground states that are more exotic than the common magnetic ones (Refs. \onlinecite{yamada:19,takimoto:08,thalmeier:19}).
\\
In the periodic  lattice the constrained fermionic mean-field treatment of the underscreened Kondo lattice model successfully captures the ingredients of the heavy quasiparticle states that form close to the Fermi level. However this approximation only involves a homogeneous global (site-independent) hybridisation field and therefore does not lead to any effective intersite couplings. The latter would appear if fluctuations of this field and their exchange between sites would be included as a next step~\cite{tesanovic:86}, but this could only capture long range interactions.  To simulate such competition effects more flexibly  even on the basis of the mean field quasiparticle picture it is customary to extend the model by adding an extra inter-site exchange term explicitly to Eq.~(\ref{eqn:cfHam}) which may be thought to have been created by having already eliminated additional higher lying conduction band states by a Schrieffer-Wolff transformation. 
This procedure has been formally carried out before in the case of the fully screened conventional KL model~\cite{riseborough:92}
. However the result has  a rather singular behaviour in \bk-space and therefore one has to resort to a phenomenological form of inter-site exchange. This leads to an extended  Kondo-Heisenberg model~\cite{irkhin:17,Bernhard:2015aa} described now by
 \be
 \bl
{\cal H}_{\rm KH}=
 &
 {\cal H}_{\rm CEF}+ \sum_{\bk\si}\epsilon_\bk c^\dag_{\bk\si}c_{\bk\si}
  \\
 &
 +
 (g_J-1)I_{ex}\sum_i\bs_i\cdot\bJ_i 
 +
 \sum_{<ij>}\bJ_i\tensor{I}_{ij}\bJ_j,
 \label{eqn:cfHeisHam}
 \el
 \ee
where $\tensor{I}$ is a cartesian uniaxial inter-site exchange tensor (counted per n.n. bond $\langle ij\rangle$) with only diagonal components $I^{x,y}_{ij}\equiv I^\perp_{ij}$ and $I^z_{ij}$. Note that for consistency we use the same sign convention for both on-site Kondo and inter-site exchange, i.e. negative for FM and positive for AF coupling. 
For reasons mentioned before we use a phenonmenological Lorentzian model for the intersite exchange of the form
\bea
I_\mu(\bq) = \frac{\Gamma^2}{\Gamma^2+(\bq-\bq_0)^2}  | I_0^\mu|,
\label{eqn:intersite}
\eea
 where $\bq_0=(0,0)$ is a zone center (FM, $I_0^\mu < 0$) or zone boundary (AF, $I^\mu_0>0$) $\bq_0=(\pi,\pi)\equiv \bQ$ wave vector, respectively and adjustable parameters $I_0^\mu, \Gamma$ characterize height and  sharpness of the maximum in $I(\bq)$ around $\bq_0$, respectively.

In RPA approximation the collective dynamical susceptibility components $(\mu=\perp,z)$ due to the last term in the above equation are then represented by
\be
\hchi_{RPA}^\mu(\bq,\nul)=
\Big[
1-I_\mu(\bq)\hchi^\mu(\bq,\nul)
\Big]^{-1}\hchi^\mu(\bq,\nul),
\label{eqn:susRPA}
\ee
where the bare magnetic susceptibilities $\hchi^\mu(\bq,\nul)$ of heavy quasiparticle bands have been  
evaluated in the previous section (Eq.~(\ref{eqn:susphys})).
The magnetic excitation spectrum of the underscreened KL as accessible by INS is then finally obtained as being proportional to the dynamical structure function $(\nul\rightarrow \omega +i0^+)$
\be
S(\bq,\omega)=\frac{1}{\pi}\frac{1}{1-e^{-\beta\omega}}
\sum_{\mu}(1-\hat{q}^2_\mu) 
\;
{\rm Im}
\Big[
\hchi^\mu_{RPA}(\bq,\omega)
\Big],
\label{eqn:struc1}
\ee
where $\hat{q}_\mu= q_\mu/|\bq|$ is normalized momentum transfer component.
%
%
\begin{figure*}
\includegraphics[width=0.99\columnwidth]{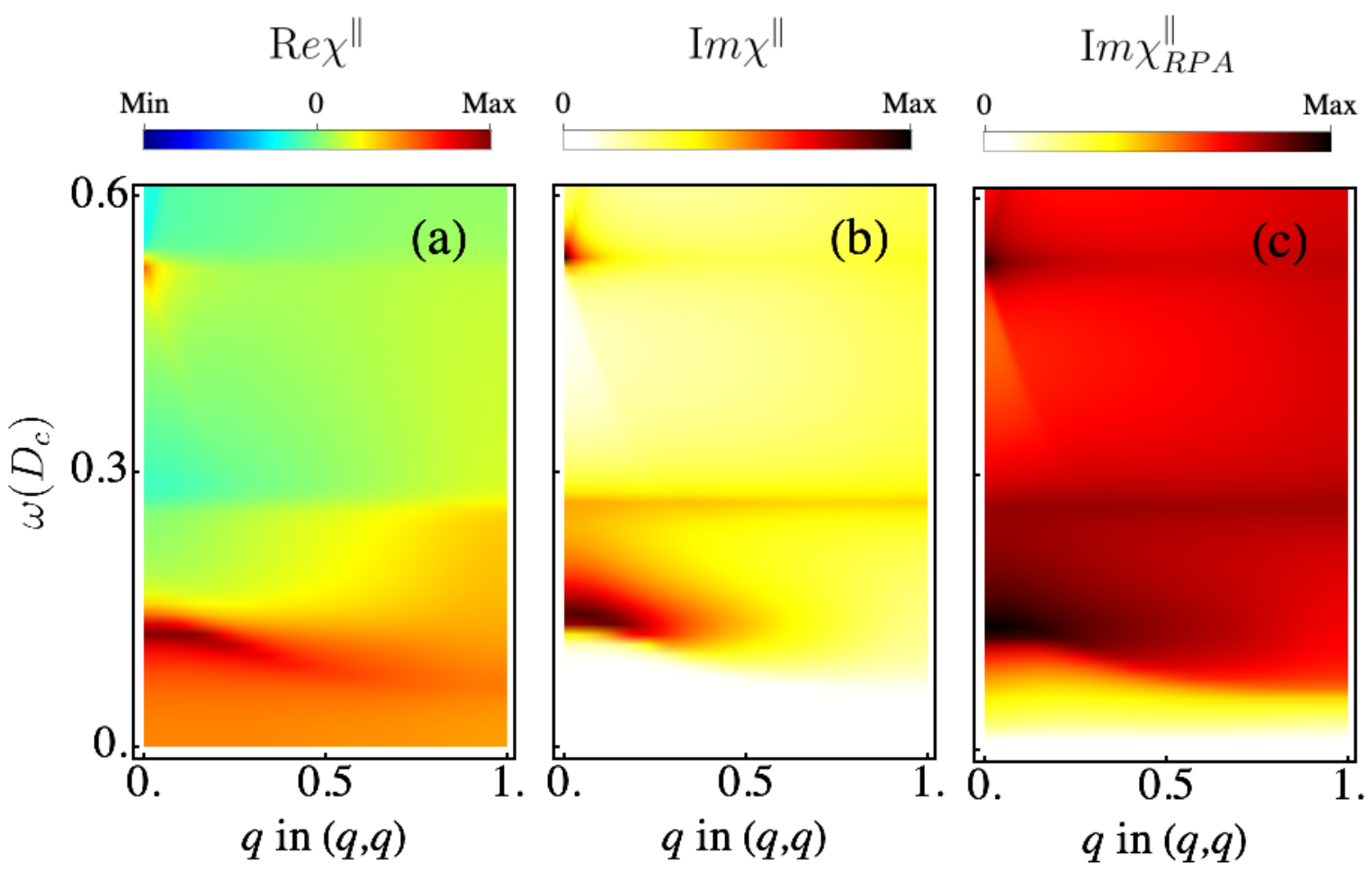}%
\hfill
%
\includegraphics[width=0.99\columnwidth]{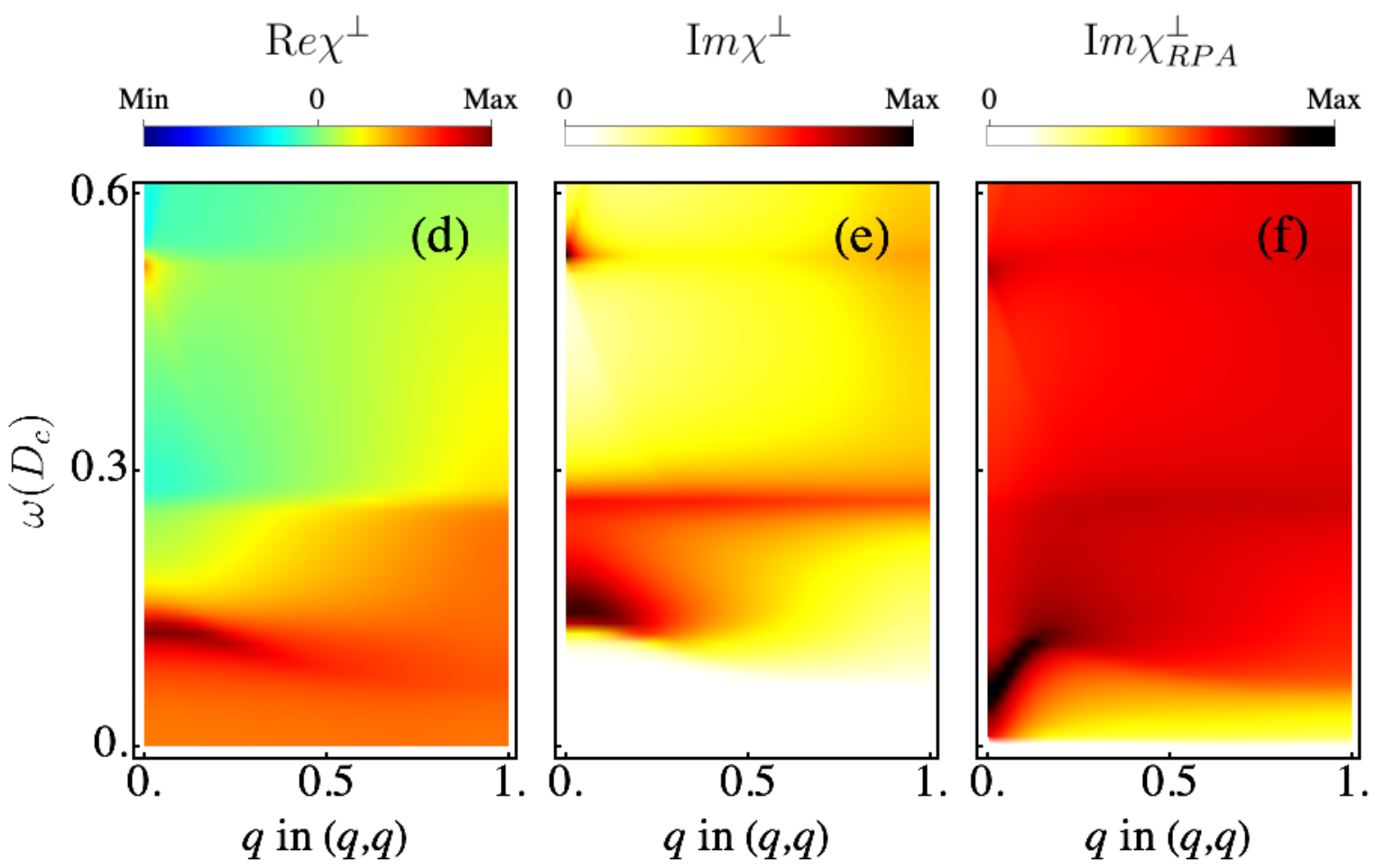}
\caption{
Frequency and momentum  dependence of the intensity  plots of  bare and RPA susceptibilities along the  $\bq=(0,0)$ to $\bq=(\pi,\pi)$ ($\Gamma$M) direction.
The left panel represents the $\parallel (J_z)$ magnetic response,  $\hchi^\parallel(\bq,T)$,
 and the second panel shows the $\perp (J_{x,y})$ magnetic response, $\hchi^\perp(\bq,T)$.
In each panel, the first and the second sub-panels indicate the real and imaginary part of the bare susceptibilities, respectively. The last sub-panels show the RPA result but in the logarithmic scale. Parameters in $I_\mu(\bq)$ same as in Fig.~\ref{fig:ChiPer}. The spectrum of  $Im\chi_{RPA}^\perp$ shows an incipient soft mode $(\bq\rightarrow 0)$ of the hybrid CEF-Kondo collective magnetic exciton.
}
\label{fig:MeshChiPer}
\end{figure*}
%

 \section{Mixed CEF-Kondo spin excitons and their temperature dependence from RPA response}
 \label{sect:RPAexciton}

We first discuss the behaviour of bare magnetic response functions in Eq.~(\ref{eqn:susphys}) which is shown in Fig.~\ref{fig:ChiPer}  for $\parallel$ (a,b) and $\perp$ directions (c,d), respectively and for zone center $\bq=(0,0)$ and zone boundary $\bq=(\pi,\pi)$ wave vectors. As in the case of optical conductivity $(\bq=0)$ one can clearly identify the two peak structure originating now from the {\it magnetic} transitions  between lower band $(n=2)$ and central $(n=3)$ as well as upper $(n=1)$ bands (Fig.~\ref{fig:BZdisp}). They are now of comparable intensity because the central band  has mainly f-electron content leading to large magnetic matrix elements. The peaks are sharper for the $\parallel$ direction whereas the spectrum (imaginary part) is more spread out for the $\perp$ direction. 
Whether they appear directly in the RPA spectrum and INS structure function $S(\bq,\omega)$ depends strongly on the type and strength of quasiparticle interactions described phenomenologically in Eq.~(\ref{eqn:intersite}).  For small $|I_0^\mu|$ the bare spectrum is hardly changed. However it is clear from Eq.~(\ref{eqn:TNRPA}) that for sufficiently large interaction when
\bea 
\frac{1}{I_\mu(\bq)}={\rm Re} \Big[ \hchi^\mu(\bq,\omega_r) \Big]
\label{eqn:spinres}
\eea
is first fulfilled for a frequency $\omega_r$ a collective magnetic resonance mode appears inside the hybridisation gaps 
($\omega_r < \Delta^d_{h1},  \Delta^d_{h3}$) [Fig.~\ref{fig:ChiPer}(e,f)] that absorbs almost all the intensity while only small features are left at the bare peak positions which are prominent in Fig.~\ref{fig:ChiPer}(a-d). Note that here the resonance is most pronounced at $\bq=0$ connected with the direct magnetic transitions.
In the conventional KL model with a single hybridization gap the bare susceptibility exhibits singular behaviour as function of frequency around the {\it indirect} gap threshold and the spin exciton resonance mode evolves at the zone boundary and inside the gap of order $T^*$ \cite{thalmeier:16}.
In contrast, in the present underscreened KL model with more realistic band structure involves both the CEF  and Kondo energy scales in direct and indirect hybridization gaps (Sec.~\ref{sec:opticalcond}) and therefore the resonance may also appear at a zone-center wave vector. This depends, however, on the precise form of $I_\mu(\bq)$ and its maximum position. The lowest hybridization gap scale is of order  $\Delta^d_{h3}=(T^{*2}+\Delta_0^2)^\fs$. Therefore a collective mode inside this gap as seen in Fig.~\ref{fig:ChiPer}(e,f) may be termed a hybrid CEF-Kondo magnetic exciton. In the limit $T^*\rightarrow 0$ it becomes the conventional CEF magnetic exciton which is the bare CEF excitation at $\Delta_0$ dispersing due to non-diagonal intersite exchange matrix elements $(\sim c_{12})$ of the bare localized two level $(\tau =1,2)$ system (Sec.~\ref{sec:induce}).\\

We also show the magnetic response and spectrum in the $(q,\omega)$ plane for \bq~along $\Gamma$M direction (Fig.~\ref{fig:MeshChiPer}). For $\parallel$ moment the bare and RPA spectrum are rather similar, meaning one is far from the resonance condition in Eq.~(\ref{eqn:spinres}). While for $\perp$ direction the comparison of bare and RPA spectrum clearly shows that a resonance mode at low energy has evolved around $\bq=0$. by properly tuning $I_\perp(\bq)$ to achieve the condition in
Eq.~(\ref{eqn:spinres}). For this direction the strength $|I^\mu_0|$ needs to be much less than for $\parallel$ direction due to the difference in the low frequency bare susceptibilities (real parts) as seen in Fig.~\ref{fig:ChiPer}.

\begin{figure}
\includegraphics[width=0.99\columnwidth]{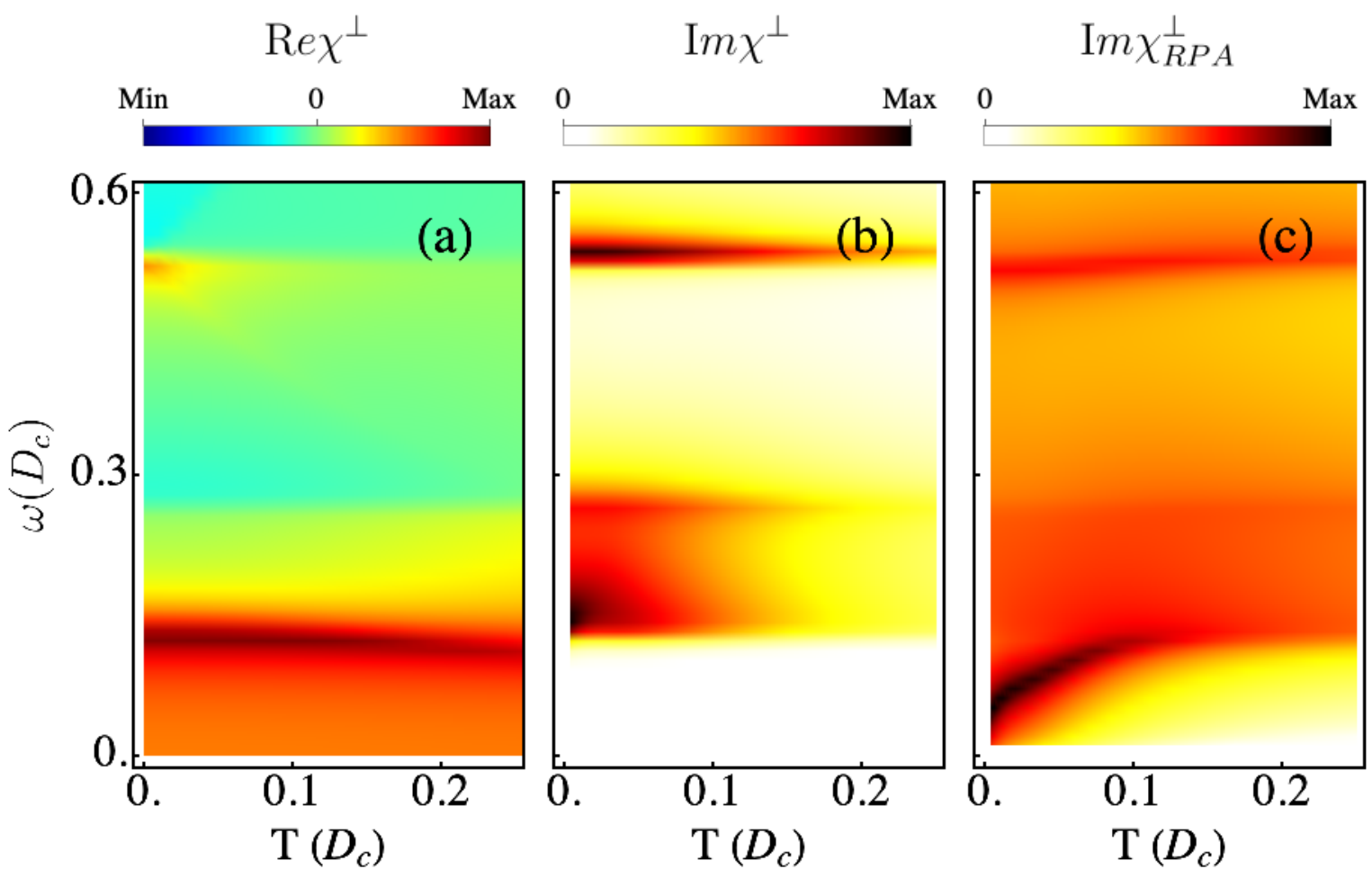}
\vspace{-0.15cm}
\caption{
Frequency and temperature  dependence of the spectral intensity  plots  for  bare and RPA 
 out of plane susceptibilities $(\perp)$ at  $\bq=(0,0)$.
The last panel (RPA) is in the logarithmic scale. It demonstrates the softening of the hybrid
CEF-Kondo magnetic exciton with decreasing temperature.
}
\label{fig:MeshChiPerT}
\end{figure}

For the parameters used the hybrid CEF-Kondo magnetic exciton in Fig.~\ref{fig:MeshChiPer}(f) shows an incipient soft-mode behaviour with $\omega_r(\bq\rightarrow 0)$ approaching zero. This is the precursor of an induced moment  FM phase transition that will appear for slightly larger coupling strength. We stress that this type of excitonic KL magnetism induced by off-diagonal exchange matrix elements  
$\sim c_{12}$ connecting different split CEF states  is fundamentally different from the usual KL magnetism \cite{zhang:00,li:10,liu:13,li:15} with fully degenerate f-states. The soft mode behaviour of the zone-center $\omega_r(\bq=0,\text T)$ is also observed as a function of temperature (Fig.~\ref{fig:MeshChiPerT}). Note that within the underlying slave-boson theory for the heavy bands the T-dependence has to be restricted to the range $kT<(T^{*2}+\Delta_0^2)^\fs$. The induced moment transition is discussed using a qualitative analytical approach in the following section.

\section{Induced magnetism criterion with Kondo screening effect}
\label{sec:induce}

In the case when the nondiagonal exchange dominates due to $c_{12}^2 \gg c^2_{\tau\tau}$ the softening of magnetic exciton mode for $T>T_c$ indicates an {\it induced} magnetic phase transition. This is well known in fully localized f-systems, e.g. in various  Pr \cite{jensen:91} and U \cite{thalmeier:02} compounds with lowest singlet-singlet CEF level scheme. The condition for the critical temperature  is obtained from the the divergence of static $(\nul=0$) RPA susceptibilities $[\mu= z,\perp(x,y)]$, i.e.
\be
\hchi_{RPA}^\mu(\bq)^{-1}=
\Big[1-I_\mu(\bq)\hchi^\mu(\bq,T)\Big]
/\hchi^\mu(\bq,T)
=0.
\label{eqn:TNRPA}
\ee
For the simple TB band structure and effective inter-site exchange model used here we can restrict to FM $(\bq =0)$ transition at $T_c$ or AF $(\bq =\bQ)$ transition at $T_N$. Naturally for the itinerant Kondo model the above equation can only be treated numerically using Eqs.~(\ref{eqn:susphys},\ref{eqn:susres}). First we recapitulate the result within the completely localized model without Kondo term but finite inter-site interaction. We consider the case $c^2_{12} \gg c^2_{\tau\tau}$ when the non-diagonal vanVleck terms dominate (Appendix \ref{sec:app2}). Then we have ($\beta=1/kT$)
\be
\hchi^\mu(\bq,T)=\frac{2m^{\mu^2}_{12}}{\De_0}\tanh\frac{\beta}{2}\Delta_0,
\label{eqn:statsus}
\ee
where we defined $m^{\mu 2}_{\tau\tau'}=\fs\sum_{\si\si'}|\langle \tau\sigma |J_\mu |\tau'\si'\rangle |^2$ and therefore 
 $2m^{\perp 2}_{12}\simeq\fs c_{12}^2$ and $m^{z2}_{12}= 0$ according to Eqs.~(\ref{eqn:pseudo},\ref{eqn:JSequiv}). The solution of Eq.~(\ref{eqn:TNRPA}) is then given by 
 \bea
 kT_{m}
 &=&
 \non
 \frac{\fs\De_0}{\tanh^{-1}\frac{1}{\xi_\bq}}
 \\
 &=&
 \left\{
\begin{array}{ll}
\xi_\bq\geq 1:& \frac{\De_0}{\ln\frac{2}{\xi_\bq-1}} \\[0.5cm]
\xi_\bq \gg 1:&  \fs\xi_\bq\De_0
\label{eqn:bandwidth3}
\end{array}
\right. ;\;\;\;  \xi_\bq=\frac{2m^{\perp 2}_{12} I_\bq^\perp}{\De_0},
\hspace{1cm}
\label{eqn:crittemp}
\eea
where $T_{m}=T_c$ for FM $(\bq=0)$ or   $T_{m}=T_N$ for AF $(\bq=\bQ)$ case, respectively. Here $\xi_\bq$ is the control parameter for induced moment magnetism which is not due to the ground state polarization alone but primarily 
($c_{12}^2 \gg c^2_{\tau\tau}$) due to the  admixiture of the excited state into the ground state by the inter-site exchange. The induced moment ground state appears only when the critical parameter fulfils $\xi_\bq >1$. This mechanism is preceded by the magnetic exciton (the bare dispersive CEF excitation) softening above $T_{cr}$. It is obtained from the pole of Eq.~(\ref{eqn:susRPA}) (for $\mu=\perp$) as
\be
\omega(\bq_0)=\De_0\bigl[1-\xi(\bq_0)\tanh\frac{\beta}{2}\Delta_0\bigr]^\fs.
\ee
This mode becomes becomes soft, i.e. $\omega(\bq_0)=0$ at the ordering temperature $T_{cr}$ and wave vector $\bq_0=0$ or  $\bq_0=\bQ$.
When the on-site Kondo coupling $I_{ex}$ to conduction electrons  is included in Eq.~(\ref{eqn:cfHam}) the localized doublets will turn into the (partly) heavy itinerant 
quasiparticle bands of Fig.~\ref{fig:BZdisp}. The excitation spectrum and critical temperature then requires the numerical 
evaluation of Eqs.~(\ref{eqn:susRPA},\ref{eqn:TNRPA}) using Eqs.~(\ref{eqn:susphys},\ref{eqn:susres}) as in the previous section. It is worthwhile, however, to have at least a qualitative understanding how the criticality condition for induced moment magnetism is modified under the presence of the Kondo screening and resulting 4f quasiparticle itineracy. This can be achieved for the FM case by using a simple analytical estimate for $\hchi^\perp(\bq=0,T)$ including only the transitions between bands $n=2,3$. This leads approximately to
\be
\bl
\hchi^\perp(0,T)=\frac{2m^{\perp^2}_{12}}{\De_{e}}\tanh\frac{\beta}{2}\Delta_0^*;
\;\;\;
\De_{e}:=\frac{\De^{*3}_0(\De^*_0-\De_0)}{\fs T^{*2}\De_0},
\\
\el
\ee
where we defined the average $\De^*_0=(T^{*2}+\De^2_0)^\fs$ with $T^*$ denoting the Kondo temperature of Eq.~(\ref{eqn:TKondo}). Furthermore  $\De_{e}$ is the effective dominating low energy scale for the vanVleck-type susceptibility contribution of transitions between occupied and empty band states $n=2,3$. Then from Eq.~(\ref{eqn:TNRPA}) we obtain the modified instability criterion for the FM case $T^*_{m}=T^*_c$ which includes the effect of Kondo screening as 
%
 %
 %
%
%
%
\begin{figure}[b]
\includegraphics[width=0.990\columnwidth]{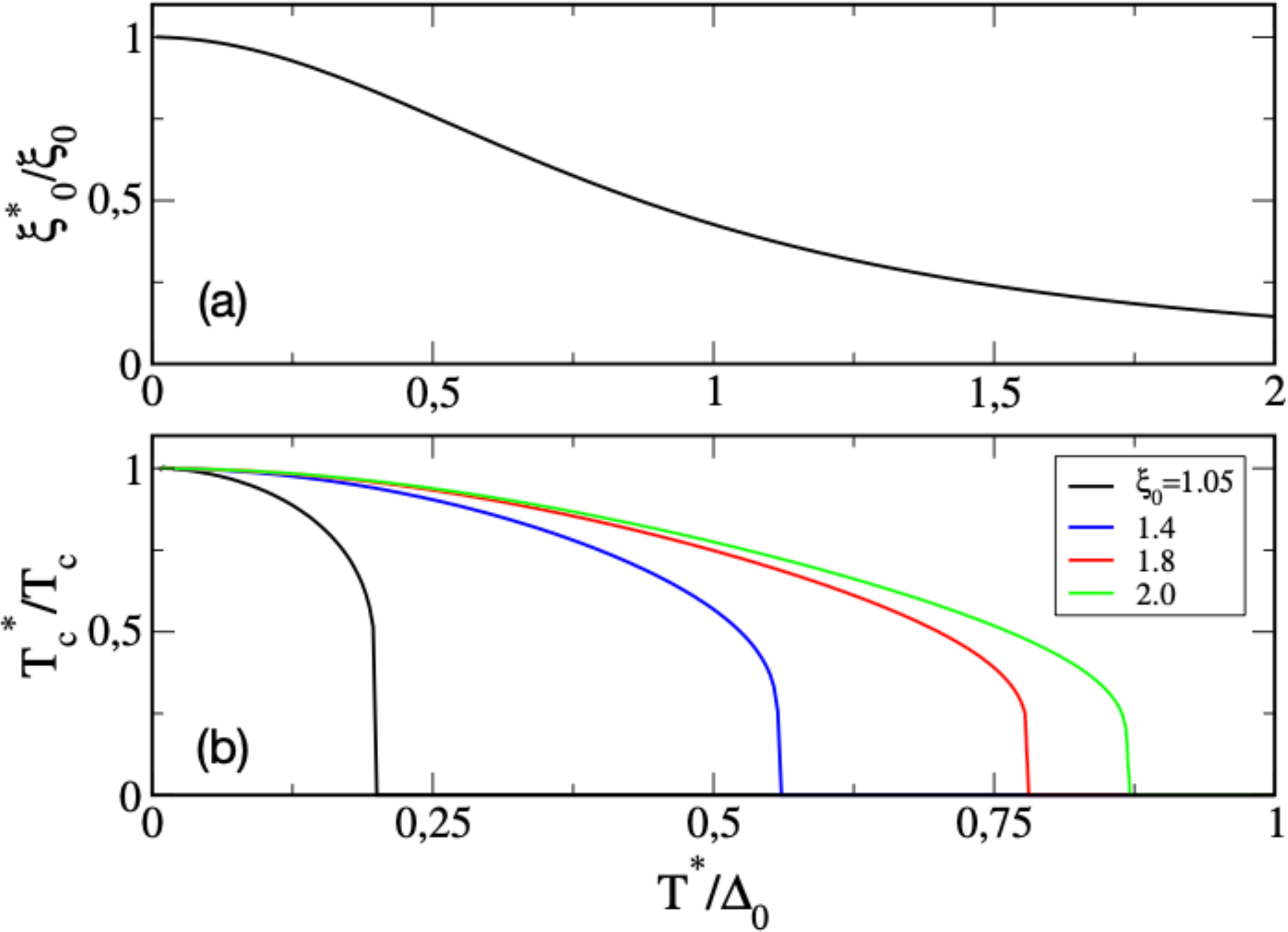}
\vspace{-0.25cm}
\caption{(top panel) Dependence of the Kondo-screened control parameter $\xi^*_0$ (normalized to the bare CEF value $\xi_0$) for induced magnetic instability on the ratio of Kondo temperature $T^*$ to CEF splitting $\De_0$. (bottom panel) Suppression of induced magnetic ordering temperature $T^*_c$  due to the strong decrease of $\xi^*_0$ (top) with increasing Kondo energy scale $T^*$, plotted for several values of bare above-critical control parameter $\xi_0 >1$. }
\label{fig:xistar}
\end{figure}
%
%
\be
kT^*_c=\frac{\fs\De^*_0}{\tanh^{-1}\frac{1}{\xi_0^*}};
\;\;\;
\xi_0^*=\frac{2m^{\perp 2}_{12} I_0^\perp}{\De_{e}},
\label{eqn:crittempFM}
\ee
where $\xi_0^*$ is the new critical parameter for induced magnetism ($\xi_0^* >1$) renormalized by the Kondo effect.
It is instructive to consider two limiting cases:
\begin{itemize}
\item{\it Nearly localized CEF excitations $T^*\ll\De_0$}:
Then $\De^*\rightarrow \De_0+\fs\frac{T^{*2}}{\De_0}$ leading to $\De_e\rightarrow \De$ and therefore $\xi_0^*\rightarrow \xi_0$ . This means in the limit of vanishing Kondo coupling $T^*\rightarrow 0$   the susceptibility $\hchi^\perp$ will be reduced to the free-ion van Vleck value of Eq.~(\ref{eqn:statsus}) for $T<T^*$. Likewise we recover the bare CEF expression in Eq.~(\ref{eqn:crittemp}) for $T_c$ in this limit.
\item{\it Dominating Kondo coupling  $T^*\gg\De_0$}:
In this case $\De^*_0\rightarrow T^*+\fs\frac{\De_0^2}{T^*}$ and therefore $\De_e\rightarrow 2T^{*2}/\De_0$. This leads to $\xi_0^*\rightarrow \fs(\De_0/T^*)^2\xi_0 \ll \xi_0$. Therefore the effective control parameter $\xi_0^*$ is much reduced and unless the bare parameter $\xi_0$ is very large the Kondo screened $\xi_0^*$ may fall below the critical value $\xi^*_{0}=1$ preventing the magnetic  instability. 
\end{itemize}
These limits imply that for all ratios of $T^*/\De_0$ we have  $\xi_0^*<\xi_0$ and the Kondo screening effect will reduce or suppress completely the appearance of the induced magnetic ordering temperature $T_{c}$.  This behaviour is illustrated in Fig.~\ref{fig:xistar}.

\section{Discussion and Conclusion}
\label{sect:discussion}

%
\begin{figure}[t]
\includegraphics[width=0.995\columnwidth]{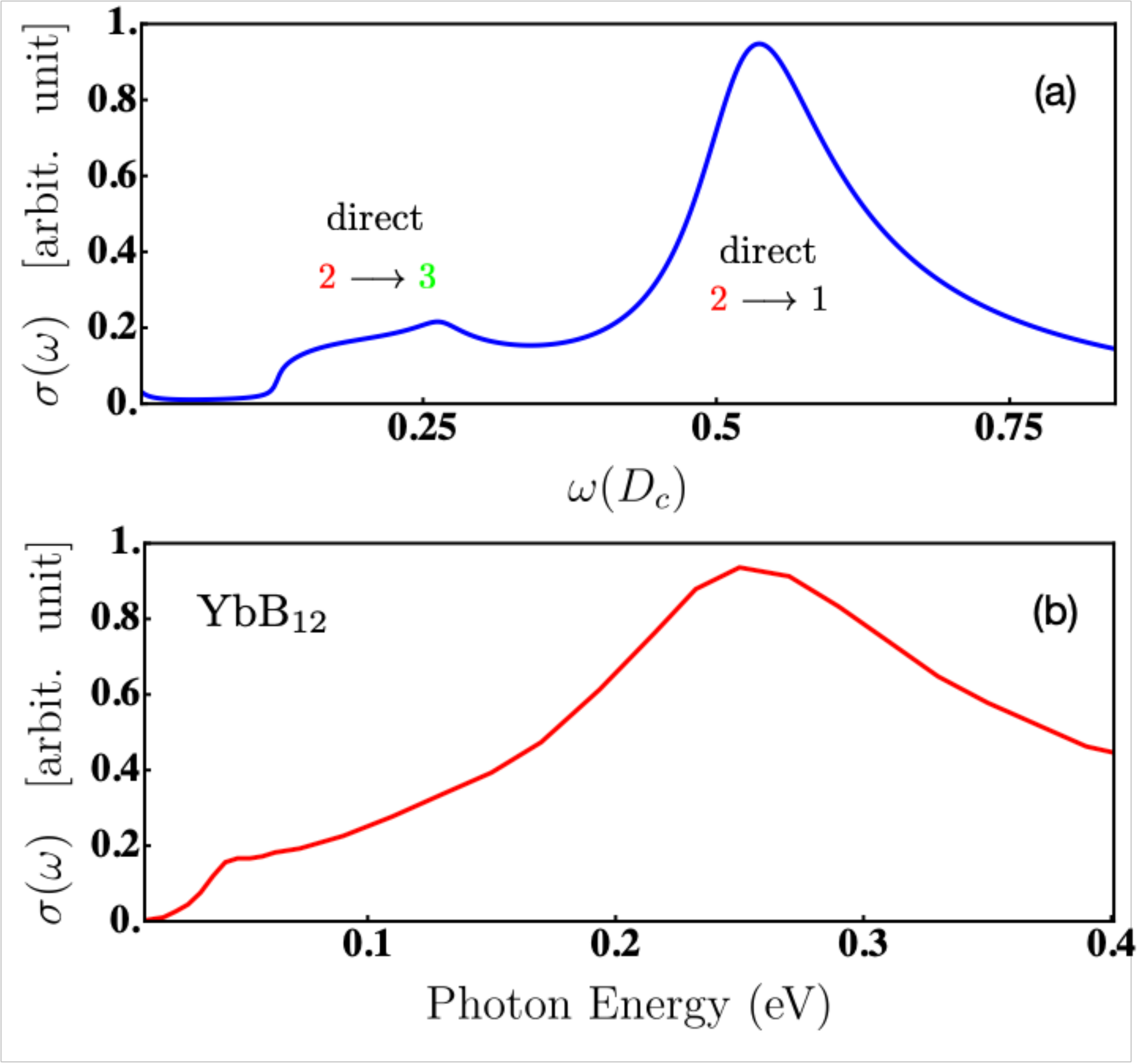}
\vspace{-0.5cm}
\caption{Qualitative comparison of optical conductivities. (a) Calculation for same parameters as Fig.~\ref{fig:optcond} but
including phenomenological linewidth \cite{mutou:94} for fermionic quasiparticles described by $\Gamma(\omega)=\gamma_0+\gamma_1\omega^2$ with $\gamma_0=0.001, \gamma_1=0.2$. Labels indicate correspondence of peak features to the {\it direct}  transitions between quasiparticle bands in Fig.~\ref{fig:BZdisp}. (b) Optical conductivity of \YB~at T$=8$ K (adapted from Ref.~\onlinecite{okamura:05}). Both direct transitions from top panel are  visible as onset shoulder and main peak. }
\label{fig:Sigma_YbB12}
\end{figure}
%

For  Kondo compounds with CEF splitting  the underscreened quasi-quartet KL model has more realistic features than the conventional SU(N) model. It shows a richer structure of quasiparticle bands around the Fermi energy that encompasses the Kondo- as well as CEF energy and effective hybridisation scales and their mutual influence. 
Since the underscreened model is realised in Ce and Yb compounds when the Kondo temperature is comparable to the
CEF splitting it is of great importance to identify these additional energy scales in inelastic experiments like finite frequency optical conductivity and 
inelastic neutron scattering. The primary goal of this work was the development of a full microscopic theory for these important probes based on the mean field slave boson solution of the model given in Ref.~\onlinecite{thalmeier:18}.\\
Firstly we found the important result that the optical conductivity which involves only direct $(\bq=0)$ transitions has a distinct two-peak structure at finite frequencies, aside from a less interesting quasielastic Drude peak. The lower peak ($\omega_l$) is dominated by the smaller scales $(T^*,\Delta)$ while the upper peak $(\omega_u)$ by the larger effective hybridisation scale $2\bav=2(T^*D_c)^\fs$ which is a non-universal scale beyond the simple Fermi liquid description.
 Only the latter is present in the conventional KL model and therefore within this model the Kondo scale $T^*$ is not directly visible in the optical conductivity. Due to the presence of the third heavy band inside the main (large) hybridization gap $2\bav$ the third heavy band leaves a direct signature in $\sigma(\omega)$ around $\omega_l$ at the Kondo/CEF scales. 

 As a corollary we briefly comment on some puzzling features found in $\sigma(\omega)$ of cubic  YbB$_{12}$~\cite{okamura:05}. There experiments not only showed a peak at the main hybridisation gap energy, as expected from the conventional KL model. It also exhibits a clear onset in $\sigma(\omega)$ at the much lower Kondo scale given by $T^*$ which in the conventional KL picture can only be associated with large momentum transitions across the {\it indirect} hybridisation gap normally not accessible for the optical response. Therefore the low-frequency onset was explained by phonon-assisted indirect transitions which are possible for zone-boundary momentum transfer carried by phonons. Our investigation suggest a possible alternative mechanism: The low frequency onset of $\sigma(\omega)$ at $T^*$ in YbB$_{12}$ can be due to {\it direct}  transitions to the central heavy band present in the underscreened KL model. In this context there would be no need to resort to indirect phonon assisted transitions.
 A qualitative comparison with the calculation of Fig.~\ref{fig:optcond}, adding a quasiparticle broadening \cite{mutou:94} for the inter-band transitions is shown in Fig.~\ref{fig:Sigma_YbB12}.
 We note however, that cubic YbB$_{12}$  has a quartet ground state and two closeby doublet excited states (i.e., a quasi-quartet split by $\Delta_0$ from the quartet ground state) excited states~\cite{alekseev:04,akbari:09}. The details of exchange-parametrization may therefore be different from the two doublet model investigated here. Further experimental evidence for multi-peak hybridization gap structure in $\sigma(\omega)$ has also been found in Ce-compounds~\cite{kimura:16}.

Secondly we demonstrated that the intricate quasiparticle band structure of the quasi-quartet KL model also shows up in the inelastic magnetic response functions probed by INS. We found that the basic ingredients of the two-peak structure due to small and large hybridization gap scales should also be present. The details depend considerably on the CEF parameters that enter as weights in the dynamic susceptibilities and on the form of the phenomenological intersite exchange.  The combined itinerant Kondo-CEF magnetic exciton spectrum may exhibit a softening as function of temperature at the wave vector where the exchange has a maximum. This is a precursor for an induced magnetic phase transition due to dominating non-diagonal exchange between the CEF-split doublets.  In the FM case $(\bq=0)$ this may be described by a simplified quasi-localized model where the control parameters a modified due to the presence of the Kondo screening. The further development of this hybrid localized-itinerant picture for CEF-Kondo magnetic excitons and induced magnetism needs an inspiration from INS and other experiments preferably on Ce- and Yb- based  Kondo lattice compounds.

\section*{ACKNOWLEDGMENTS}
A.A. acknowledges financial support from the National Research Foundation (NRF) funded by the Ministry of Science of Korea (Grants: No. 2016K1A4A01922028, No. 2017R1D1A1B03033465, and No. 2019R1H1A2039733).

\appendix

\section{}
\label{sec:app1}

In this appendix we give the explicit expressions of spectral residua entering the pseudospin susceptibilities in Eq.~(\ref{eqn:susres}) as well as those in the optical conductivity of Eq.~(\ref{eqn:siginter}). 
For the former $\hchi_{\tau\tau}(\bq,\nul)$ and  $\hchi^a_{\tau\bartau}(\bq,\nul)$ we have:
\be
\bl
\hZ^\tau_{1\bk}=
&\frac{\bav^2_\tau |E_{1\bk}-\eplab|}{|E_{1\bk}-E_{2\bk}||E_{1\bk}-E_{3\bk}||E_{1\bk}-\epla |},
\\
\hZ^\tau_{2\bk}=&\frac{\bav^2_\tau |E_{2\bk}-\eplab|}{|E_{2\bk}-E_{1\bk}||E_{2\bk}-E_{3\bk}||E_{2\bk}-\epla |},
\\
\hZ^\tau_{3\bk}=&\frac{\bav^2_\tau |E_{3\bk}-\eplab|}{|E_{3\bk}-E_{1\bk}||E_{3\bk}-E_{2\bk}||E_{3\bk}-\epla |}.
\el
\label{eqn:hatres}
\ee
Likewise the spectral residua for the orbitally nondiagonal contribution   $\hchi^b_{\tau\bartau}(\bq,\nul)$ are given by
\be
\bl
\hZ^B_{1\bk}=&\frac{\bav_1\bav_2\si_{1\bk}}{|E_{1\bk}-E_{2\bk}||E_{1\bk}-E_{3\bk}|},
\\
\hZ^B_{2\bk}=&\frac{\bav_1\bav_2\si_{2\bk}}{|E_{2\bk}-E_{1\bk}||E_{2\bk}-E_{3\bk}|},
\\
\hZ^B_{3\bk}=&\frac{\bav_1\bav_2\si_{3\bk}}{|E_{3\bk}-E_{1\bk}||E_{3\bk}-E_{2\bk}|},
\el
\label{eqn:hatresB}
\ee
with the sign 
$\si_{\bt\bk}=\pm 1$, defined in Eq.~(\ref{sign}).
%

In the case of the optical $\bq=0$ conductivity $\sigma(\omega)$ in  Eq.~(\ref{eqn:siginter}) we need the residua
\be
\bl
\tiZ_{1\bk}=&\frac{|E_{1\bk}-\eplax| |E_{1\bk}-\eplay|}{|E_{1\bk}-E_{2\bk}||E_{1\bk}-E_{3\bk}|},
\\
\tiZ_{2\bk}=&\frac{|E_{2\bk}-\eplax| |E_{2\bk}-\eplay|}{|E_{2\bk}-E_{1\bk}||E_{2\bk}-E_{3\bk}|},
\\
\tiZ_{3\bk}=&\frac{|E_{3\bk}-\eplax| |E_{3\bk}-\eplay|}{|E_{3\bk}-E_{1\bk}||E_{3\bk}-E_{2\bk}|}.
\el
\label{eqn:tilres}
\ee

\section{}
\label{sec:app2}
%
\begin{figure}[t]
\includegraphics[width=0.995\columnwidth]{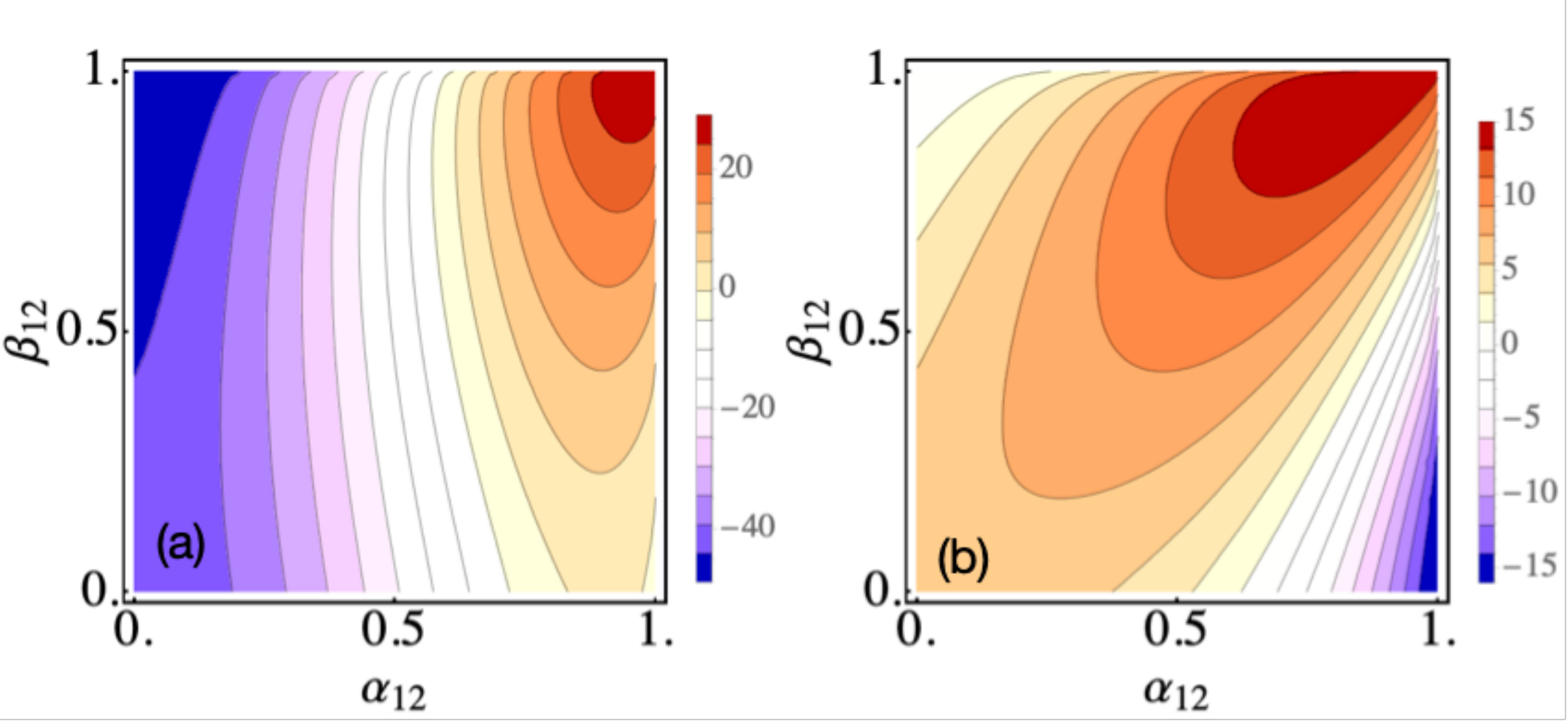}
\vspace{-0.5cm}
\caption{Contour plot of the differences of squared moment coefficients  $c_{12}^2-c^{z2}_{11}$ (left) and  $c_{12}^2-c^{2}_{11}$ (right) for ground state doublet ($\tau =1$) in the $\alpha_{12}-\beta_{12}$ plane of CEF state parameters. In the upper right corner where  $\alpha_{12},\beta_{12}\rightarrow 1$ the nondiagonal coefficents dominate, i.e.  $c_{12}^2\gg c^{z2}_{11}, c^{2}_{11}$. For excited state $\tau =2$ the behaviour is qualitatively similar.
 }
\label{fig:momentcoeff}
\end{figure}
%

\begin{table}
\begin{center}
\begin{tabular}{|c|c|c||c|c||c|c||c|c|c|}
 \hline
 $J^\perp_1$ & $ J^\perp_2 $ & $J_{12}$ &
$ \alpha_{12}$ & $ \beta_{12}$ & $c^z_{11} $ & $c^z_{22}$ & $c_{11}$ & $c^{}_{22}$ & $c^{}_{12}$
 \\
 \hline
%
0.471
&
0.767
&
0.2
&
 0.59 & 
 0.35 & 
 4.18 & 
 -4.0 & 
 1.4 &
  2.3 & 
0.84
 \\
 \hline
 \end{tabular}
 \caption{The values of the  the anisotropy coefficients obtained based on original model parameters ($J^\perp_1$, $ J^\perp_2 $,  $J_{12}$). The energy scale is $D_c$. }
\label{table:1}
\end{center}
\end{table}

Here we briefly discuss the origin of the anisotropy coefficients  $c^z_{\tau\tau'}$ and $c_{\tau\tau'}$ that are essential for  the relation between total angular momentum $\bJ$ and pseudo spin $\bS$ [Eq.~(\ref{eqn:JSequiv})] and enter as well  the physical (cartesian) bare susceptibilities in Eq.~(\ref{eqn:susphys}). The coefficients are determined by the composition of quasi-quartet CEF states consisting of a $\Gamma_6$-$\Gamma_7$ pair given by
\be
\bl
(\tau=1):\;\;\; |\Ga_6\pm\ra&=\alpha_{11}|\pm\frac{7}{2}\ket+\alpha_{12}|\mp\fs\ket,
\\
(\tau=2):\;\;\; |\Ga_7\pm\ra&=\beta_{11}|\mp\frac{5}{2}\ket+\beta_{12}|\pm\frac{3}{2}\ket,
\el
\label{eqn:CEFstate}
\ee
Using the normalization conditions $\alpha_{11}^2+\alpha_{12}^2=1$ and $\beta_{11}^2+\beta_{12}^2=1$ 
the moment coefficients may be obtained as function of independent CEF parameters $\alpha_{12}$ and $\beta_{12}$
\cite{takimoto:08}:
\be
\bl
c^z_{11}&=7-8\alpha^2_{12}; \;\;\;  c^z_{22}=-5+8\beta^2_{12},
\\
c_{11}&=4\alpha_{12}^2; \;\;\; c_{22}=4\sqrt{3}\beta_{12}\sqrt{1-\beta^2_{12}},
\\
c_{12}&=\sqrt{7(1-\alpha^2_{12})(1-\beta^2_{12})} +\sqrt{30}\alpha_{12}\beta_{12}.
\label{eqn:Jcoeff}
\el
\ee
The relative size of these coefficients, characterized e.g. by their differences $c_{12}^2-c^{z2}_{\tau\tau}$  and 
 $c_{12}^2-c^2_{\tau\tau}$ ($\tau=1,2$) varies greatly with CEF state parameters $\alpha_{12},\beta_{12}$. The most interesting case is the `induced-moment' situation when off diagonal coefficients between the two doublets are dominating the magnetic response and possible ordering, i.e. $c_{12}^2\gg c^{z2}_{\tau\tau},c^{2}_{\tau\tau}$. The Fig.~\ref{fig:momentcoeff} which plots the above differences  in the $\alpha_{12},\beta_{12}$ plane shows that this situation can be reached in the upper right corner where when $\alpha_{12},\beta_{12}\rightarrow 1$ meaning $\alpha_{11}, \beta_{11}\rightarrow 0$. This corresponds to CEF doublets in Eq.~(\ref{eqn:CEFstate}) dominated by $|\mp\fs\ket$ and $|\pm\frac{3}{2}\ket$ states which  may be the case when the parameters of the tetragonal CEF satisfy $|B_2^0| \gg |B^m_{4}|,  |B^m_{6}|$.
Then the moment operators in Eq.~(\ref{eqn:JSequiv}) are essentially of easy-plane type $J_z\approx 0$ and $J_\pm\approx c_{12}\frac{1}{\sqrt{2}}(S^\pm_{12}+S^\pm_{21})$ where the latter has only contributions {\it off-diagonal} (12, 21) in the CEF doublet states. This corresponds to dominating inelastic magnetic response due to the quasi-quartet CEF splitting and a possible excitonic magnetic order with primarily induced moments due to the mixing of $\tau=1,2$ by inter-site exchange.

\bibliography{References}

\end{document}